\documentclass{article} 
\usepackage{iclr2026_conference,times}

\usepackage{amsmath,amsfonts,bm}

\def\eqref#1{equation~\ref{#1}}

\def\1{\bm{1}}

\DeclareMathAlphabet{\mathsfit}{\encodingdefault}{\sfdefault}{m}{sl}
\SetMathAlphabet{\mathsfit}{bold}{\encodingdefault}{\sfdefault}{bx}{n}

\usepackage{hyperref}
\usepackage{url}
\usepackage{graphicx} 
\usepackage{wrapfig}
\usepackage{multirow}
\usepackage{enumitem}
\usepackage[normalem]{ulem}
\usepackage[most]{tcolorbox}
\usepackage{colortbl}
\usepackage{geometry}

\usepackage{booktabs}
\usepackage{tabularx}
\usepackage{array}
\newcolumntype{L}{>{\raggedright\arraybackslash}X}
\newcolumntype{R}{>{\raggedleft\arraybackslash}X}
\usepackage{graphicx}
\usepackage{subcaption}  
\usepackage{caption}

\usepackage{listings}
\usepackage{xcolor}

\usepackage{bbm}
\usepackage{amssymb}

\definecolor{codebg}{RGB}{245, 245, 220}  

\lstdefinestyle{whitebg}{
    backgroundcolor=\color{white},
    basicstyle=\ttfamily\small,
    breaklines=true,
    frame=single,
    framesep=10pt,
    xleftmargin=10pt,
    xrightmargin=10pt,
    framexleftmargin=10pt,
    framexrightmargin=10pt,
    rulecolor=\color{gray},  
}

\title{Computer-Using World Model}

\author{
Yiming Guan$^{1,4*\,\dagger}$ \quad
Rui Yu$^{2,4*\,\dagger}$ \quad
John Zhang$^{3,4*\dagger}$ \quad
Lu Wang$^{4\ddagger}$ \quad
Chaoyun Zhang$^{4}$ \\
Liqun Li$^{4}$ \quad
Bo Qiao$^{4}$ \quad
Si Qin$^{4}$ \quad
He Huang$^{4}$ \quad
Fangkai Yang$^{4}$ \\
Pu Zhao$^{4}$ \quad
Lukas Wutschitz$^{4}$ \quad
Samuel Kessler$^{4}$ \quad
Huseyin A. Inan$^{4}$ \quad
Robert Sim$^{4}$ \\
Saravan Rajmohan$^{4}$ \quad
Qingwei Lin$^{4}$ \quad
Dongmei Zhang$^{4}$ \\[6pt]
$^{1}$Nankai University \quad
$^{2}$Nanjing University \quad
$^{3}$The University of New South Wales \quad
$^{4}$Microsoft \\
\small{$^*$Equal contribution \quad
$^\dagger$Microsoft intern \quad
$^\ddagger$Corresponding author}
}

\iclrfinalcopy 
\begin{document}

\maketitle
\begin{abstract}
Agents operating in complex software environments benefit from reasoning about the consequences of their actions, as even a single incorrect user interface (UI) operation can derail long, artifact-preserving workflows. This challenge is particularly acute for computer-using scenarios, where real execution does not support counterfactual exploration, making large-scale trial-and-error learning and planning impractical despite the environment being fully digital and deterministic. We introduce the Computer-Using World Model (CUWM), a world model for desktop software that predicts the next UI state given the current state and a candidate action. CUWM adopts a two-stage factorization of UI dynamics: it first predicts a textual description of agent-relevant state changes, and then realizes these changes visually to synthesize the next screenshot. CUWM is trained on offline UI transitions collected from agents interacting with real Microsoft Office applications, and further refined with a lightweight reinforcement learning stage that aligns textual transition predictions with the structural requirements of computer-using environments. We evaluate CUWM via test-time action search, where a frozen agent uses the world model to simulate and compare candidate actions before execution.
Across a range of Office tasks, world-model-guided test-time scaling improves decision quality and execution robustness.

\end{abstract}

\section{Introduction}

The performance of Large language models (LLMs) has improved consistently through scaling; natural language has been effectively modeled by training large models on massive static corpora~\citep{brown2020language,hoffmann2022training,openai2024gpt4technicalreport}. In contrast, agents operate in non-stationary environments: their actions shape future observations and thus the data they learn from changes as learning progresses~\citep{sutton1999policy,sutton1998reinforcement,yao2022react,yao2023tree}. Since an agent’s actions change the world’s state, reliable decision-making requires counterfactual reasoning: anticipating the consequences of alternative actions before choosing one.

This requirement is particularly acute for \emph{computer-using} agents in desktop applications~\citep{zhang2025ufo,zhang2025ufo2,zhang2025ufo3,bonatti2024windows}. Although software is fully digital and largely deterministic, interaction is neither cheap nor safely reversible: UI actions incur substantial latency~\citep{endo1996using}, undo is limited and context-dependent~\citep{prakash1994framework}, and a single mistake can corrupt artifacts or derail long workflows. Determinism therefore does not imply cheap rollouts; without simulation, desktop agents cannot effectively and safely perform counterfactual explorations, making trial-and-error learning and real-execution tree search impractical~\citep{xie2024osworld,zhou2023webarena,chen2025scaling}.

While model-based reinforcement learning has demonstrated the value of learned dynamics in robotics and games~\citep{ha2018world, levine2022understanding, schrittwieser2020mastering, hafner2019learning}, world models remain underexplored for \emph{GUI-based desktop software} in the era of LLMs \citep{zhang2024large}.
Recent LLM-based world modeling efforts have primarily focused on implicit latent dynamics, textual or semantic state transitions for web and mobile agents, or visual observation prediction in mobile UI settings~\citep{hafner2019learning, hafner2023mastering, chae2world, li2025mobileworldbench, li2025word, luo2025vimo, cao2026mobiledreamer, xiang2025uisim}, rather than interactive desktop GUIs.
Desktop software for computer use poses unique challenges, combining high-dimensional visual observations, rich compositional GUI actions, and long-horizon, artifact-preserving workflows where early mistakes persist and compound \cite{wang2024large}.

In this paper, we take a first step toward world modeling for computer use by introducing the \emph{Computer-Using World Model} (CUWM) for real-world desktop software.
We instantiate CUWM in the Microsoft Office suite, including \emph{Word}, \emph{Excel}, and \emph{PowerPoint}, which are widely used productivity applications.
World modeling in this domain is especially important because actions such as editing, formatting, or deleting content can have irreversible consequences; a faithful world model enables agents to simulate action outcomes for safer planning, faster evaluation, and more reliable automation without interacting with live user data.
\begin{wrapfigure}{r}{0.44\textwidth}
  \centering
  \vspace{-15pt}
  \includegraphics[width=0.44\textwidth]{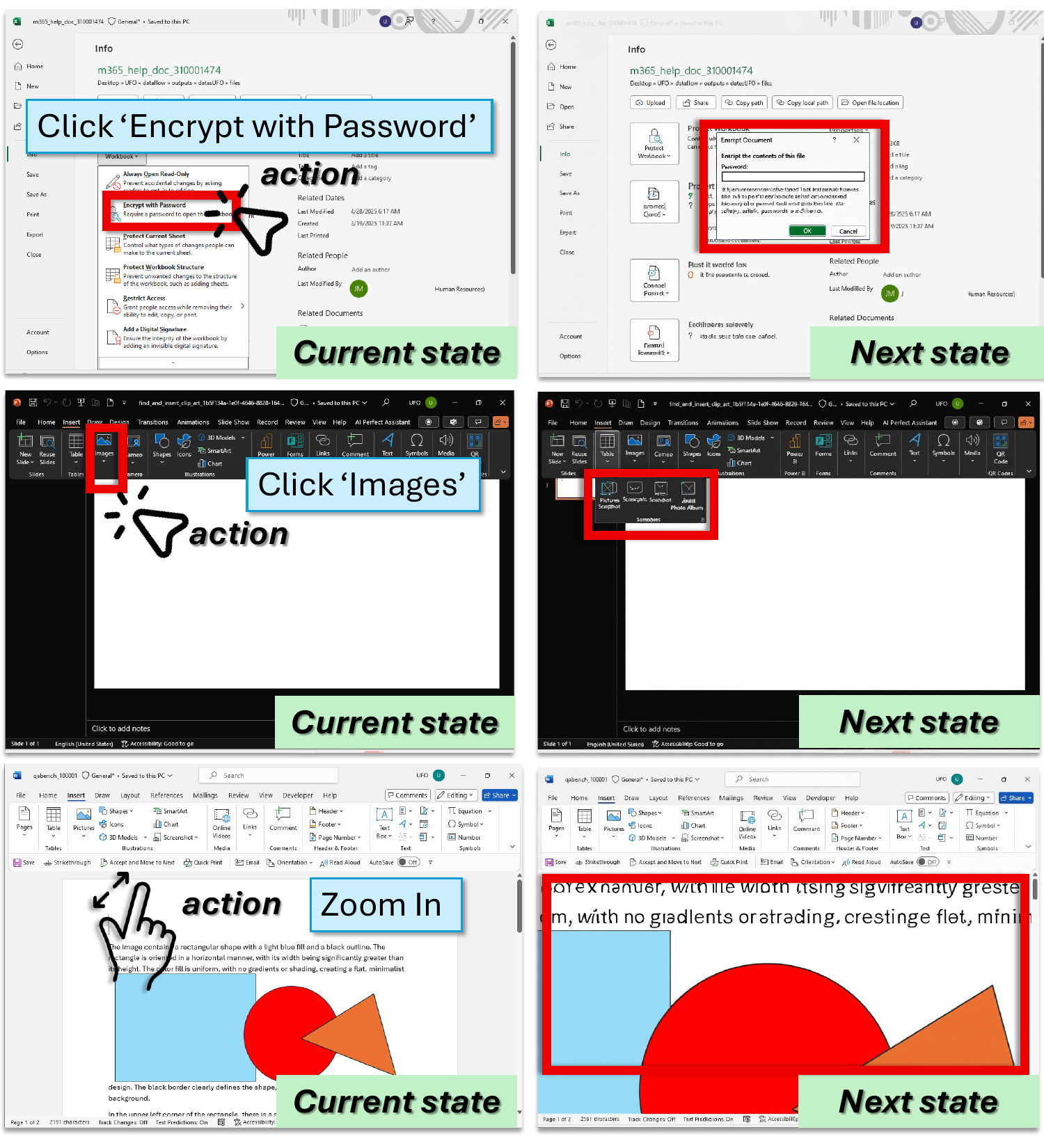}
  \vspace{-15pt}
\caption{UI state transitions generated by CUWM. Each row is one example transition.}
  \label{fig:case_intro}
\vspace{-15pt}
\end{wrapfigure}

CUWM predicts the next UI state from the current state and a candidate action by factorizing UI dynamics into two stages.
A textual transition model first predicts the action-induced, decision-relevant UI changes, and a visual realization model then renders these changes as the next screenshot.
This separation of \emph{what changes} from \emph{how it appears} focuses model capacity on structurally salient transitions while retaining pixel-level state generation required by desktop agents.
Figure~\ref{fig:case_intro} shows example UI state transitions predicted by CUWM, where candidate actions induce localized but consequential interface changes.
CUWM is trained on offline UI transitions collected from agents interacting with real Microsoft Office applications.
Supervised fine-tuning provides a faithful initialization of UI dynamics, which is further refined with a lightweight reinforcement learning stage that uses an LLM-based judge and a length penalty to encourage concise textual transitions that are aligned with the structural organization of software UIs.

We evaluate CUWM using \emph{test-time action search} with a frozen LLM agent, where candidate actions are simulated by CUWM and a single action is executed.
This enables improved decision quality through additional test-time computation without further training or risky exploration, and we evaluate both agent performance and the fidelity of predicted textual transitions and next-state screenshots. In summary, this work makes the following:
\vspace{-0.5em}
\begin{itemize}[leftmargin=*]
    \item To our best knowledge, \textbf{we present the first Computer-Using World Model (CUWM) that explicitly models UI state transitions}, enabling test-time planning for productivity use-cases by learning to reconstruct Microsoft Office (Word, Excel, and PowerPoint) applications.
\item \textbf{We propose a two-stage world model learning framework} that factorizes UI state transitions into a \emph{textual abstraction} of action-induced changes followed by a \emph{visual state realization}.
CUWM is initialized via supervised learning on offline UI transitions and further refined with a reinforcement learning stage that aligns textual transitions with the structural organization of software UIs, promoting concise descriptions.
\item Taken together, our world model evaluations show that test-time simulation of UI consequences can substantially improve reliable decision making in software systems.

\end{itemize}

\vspace{-1em}
\section{Related Work}
\textbf{Implicit World Models.}
Model-based reinforcement learning has long studied \emph{implicit} world models~\citep{ha2018world,levine2022understanding, schrittwieser2020mastering,hafner2019learning} that encode environment dynamics in latent representations for planning and value prediction, rather than explicit state reconstruction.
Representative methods include World Models~\citep{ha2018world}, PlaNet~\citep{hafner2019learning}, Dreamer~\citep{hafner2023mastering}, and MuZero~\citep{schrittwieser2020mastering}.
While effective in games and robotics, these latent models are not designed to be interpretable or aligned with explicit UI semantics, limiting their applicability to computer-using agents.
\textbf{Textual and Semantic World Models.}
Recent work explores explicit textual or semantic world models, primarily for web and mobile agents.
Web Agents with World Models~\citep{chae2world} predicts semantic state transitions for test-time action search in web navigation, and MobileWorldBench~\citep{li2025mobileworldbench} argues for the effectiveness of semantic prediction in mobile environments.
Related studies such as ``From Word to World"~\citep{li2025word} examine whether large language models can serve as implicit world models in purely text-based settings.
However, these approaches do not model the visual realization of UI changes, which is often critical for desktop software.
\textbf{Visual GUI World Models.}
Another line of work focuses on predicting future GUI observations.
ViMo~\citep{luo2025vimo} and MobileDreamer~\citep{cao2026mobiledreamer} synthesize future app screens using visual world models, while UI-SIM~\citep{xiang2025uisim} explores image-based UI simulation.
In contrast, we present a factorized, multimodal world model for desktop productivity software that, to the best of our knowledge, is the first world model explicitly tailored for \emph{GUI-based computer use} in software environments, where small UI changes can have outsized effects in long, artifact-preserving workflows.

\begin{figure}[t]
  \centering
  \includegraphics[width=0.85\linewidth]{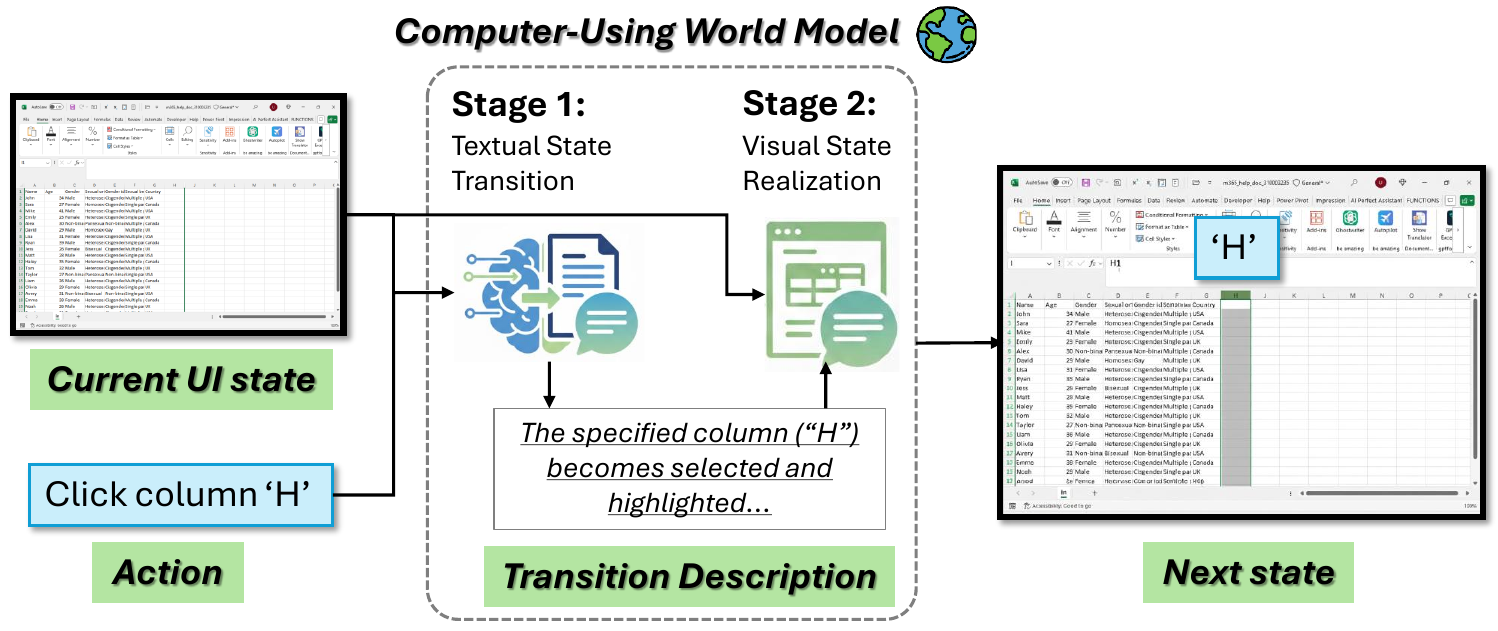}
  \caption{\textbf{Overview of the CUWM.} The world model state transitions proceed in two stages, in the first stage, given the current UI state and an action, the world model predicts a textual state-transition description of the next state. In the second stage, the world model conditions on the current UI state and the transition description to render the next UI state.}
  \label{fig:cuwm_overview}
\end{figure}
\vspace{-12pt}
\section{Method}
\label{sec:method}

Our goal is to learn a world model that captures software UI dynamics and supports agent decision-making through imagined trajectories.
Given a current UI state and a candidate action, the model predicts the resulting next state using annotated UI transitions collected from real applications.
We instantiate this objective in desktop productivity software and propose a two-stage \emph{computer-using world model}, as illustrated in Figure~\ref{fig:cuwm_overview}.
The model factorizes UI dynamics into two stages.
Stage~1 (\emph{Textual State Transition}) predicts a structured natural-language description of the localized, decision-relevant change induced by the action.
Stage~2 (\emph{Visual State Realization}) conditions on the current UI state and the predicted transition to synthesize the next UI screenshot, preserving unchanged regions while applying the specified edits.
This factorization separates \emph{what changes} from \emph{how it appears}, enabling interpretable and controllable modeling of UI dynamics.
The world model is trained primarily via supervised learning on offline UI transitions, and further refined with reinforcement learning to improve the decision relevance and conciseness of predicted transitions.

\subsection{Two-Stage World Model Architecture}
\label{sec:architecture}
\vspace{-7pt}

We model computer-using interaction as a sequential decision process and instantiate our study in desktop productivity software.
At time step $t$, the environment is in a UI state $s_t$ represented by a screenshot image, and executing a natural-language action $a_t$ induces a transition to the next UI state $s_{t+1}$.

Directly predicting the next UI state $s_{t+1}$ from $(s_t, a_t)$ in pixel space is computationally inefficient due to the highly structured and sparse nature of software state changes.
\textbf{We define this structure by three key characteristics: UI transitions are typically \textit{localized} in space, \textit{compositional} in nature, and \textit{causally aligned} with the triggering action.}
In desktop applications, most actions induce \emph{localized} updates, such as changing a selection, spawning a dialog box, or moving a text cursor, while the majority of the interface remains unchanged.
While this sparsity is rooted in a strong underlying structure, it complicates monolithic pixel-level prediction by forcing models to simultaneously track massive invariant backgrounds and tiny, decision-critical updates.
Consequently, end-to-end pixel prediction often wastes modeling capacity on these static regions and fails to emphasize the action-relevant components of the transition.

To exploit this structure, we adopt a two-stage decomposition of software UI dynamics.
Specifically, we separate \emph{what changes} from \emph{how it appears} by first predicting a textual abstraction of the action-induced state transition, followed by a visual realization of the next UI state.
This factorization allows the model to explicitly represent semantically meaningful changes, such as which UI element is affected and in what manner, while delegating image synthesis to a specialized visual model.
By aligning the model architecture with the compositional and localized nature of UI interactions, the world model can focus its capacity on decision-relevant dynamics rather than static visual details.

\textbf{Stage 1: Textual State Transition Model.}
This stage predicts a textual description of the UI transition. We employ Qwen2.5-VL~\citep{bai2025qwen25vltechnicalreport} as a vision-language model that takes the current UI state $s_t$ (including the screenshot and associated textual context) and the action $a_t$ as input, and outputs a textual transition description $\Delta_t$:
$
\Delta_t = f_{\text{text}}(s_t, a_t).
$
Rather than describing the entire UI, $\Delta_t$ focuses on decision-relevant changes, such as selection shifts, content edits, dialog appearances, or mode transitions. This abstraction significantly reduces the prediction space and provides an interpretable representation of software dynamics that is well aligned with agent decision-making.

\textbf{Stage 2: Visual State Realization Model.}
The second stage translates the abstract transition description into a concrete visual outcome.
We use Qwen-Image-Edit\footnote{We use the Qwen-Image-Edit-2509 checkpoint throughout this work.}~\citep{wu2025qwenimagetechnicalreport}, a diffusion-based conditional image editing model, to synthesize the next-state screenshot conditioned on the current UI and the predicted transition:
$
\hat{s}_{t+1} = f_{\text{image}}(s_t, \Delta_t).
$
By conditioning on both the current screenshot and the textual transition, the visual model is responsible only for rendering localized changes, while preserving unchanged regions.

This two-stage design cleanly separates semantic state transitions from visual realization.
By allocating model capacity to the most informative components of the prediction problem other than pixel-level synthesis, the architecture improves interpretability, modularity, and scalability, and forms the basis for effective training and downstream agent usage.

\subsection{Supervised Training with GPT-Annotated Transitions}
\label{sec:sft}

We first initialize the \emph{Computer-Using World Model} (CUWM) using supervised learning.
Supervised training provides a natural starting point for learning faithful software dynamics, as it allows the model to directly observe how UI states change in response to actions.
However, manually annotating UI transitions at scale is prohibitively expensive, motivating the use of automated supervision.

To obtain training data, we use the GUI-360~\citep{mu2025gui} dataset, which consists of UI interaction trajectories generated by multiple computer-using agents interacting with Office applications.
Each trajectory is represented as a sequence of screenshot-based UI states and text-described GUI/API actions, yielding transition tuples $(s_t, a_t, s_{t+1})$, as illustrated in Figure~\ref{fig:case_intro}.
Based on these transitions, we use GPT-5 as an automated annotator to generate concise natural-language description of UI state changes by conditioning on the triplet $(s_t, a_t, s_{t+1})$, explicitly identifying which elements change and which remain unchanged.

This process yields \emph{ground-truth} transition descriptions of the form $(s_t, a_t) \rightarrow \Delta_t^{\text{GT}},$
where $\Delta_t^{\text{GT}}$ summarizes the semantic differences between the consecutive UI states $s_t$ and $s_{t+1}$.
We apply supervised fine-tuning to both stages of CUWM.
In Stage~1, the textual transition model (Qwen-VL~2.5) is trained to predict the GPT-annotated transition description $\Delta_t^{\text{GT}}$ from the current screenshot and action $(s_t, a_t)$, producing a predicted transition $\Delta_t$.
In Stage~2, the visual realization model (Qwen-Image-Edit) is trained via diffusion-based image editing to reconstruct the next UI state $s_{t+1}$ conditioned on the current screenshot and the predicted transition $(s_t, \Delta_t)$.
This supervised training grounds both stages in real software behavior and provides a faithful initialization of UI dynamics and a strong foundation for subsequent refinement.  

\vspace{-12pt}
\subsection{Structure-Aware Reinforcement Learning for Textual Transitions}
Supervised fine-tuning provides a faithful initialization of textual UI transitions, but it does not ensure that the predicted descriptions consistently capture the UI structures most critical for downstream reasoning.
In computer-using environments, effective planning depends on whether key interface components, such as selection state, active controls, and visible panes, are accurately and concisely represented.

We therefore apply a lightweight reinforcement learning refinement to the textual transition model.
The model is treated as a policy conditioned on the current screenshot and action $(s_t, a_t)$, generating a transition description $\Delta_t$, and is optimized to maximize
$\mathbb{E}_{\Delta_t \sim f_{\text{text}}(\cdot \mid s_t, a_t)} 
\left[ R(s_t, a_t, \Delta_t) \right].
$
The reward combines an LLM-as-a-Judge score and a length penalty:
$R(s_t, a_t, \Delta_t)
= R_{\text{judge}}(\Delta_t, \Delta_t^{\mathrm{GT}}) - \beta R_{\text{len}}(\Delta_t).$
The judge assigns a normalized score in $[0,1]$ by evaluating correctness across predefined UI structural aspects (e.g., ribbon state, editing area, and side panes).
The length penalty softly penalizes predictions that deviate from a target length range defined relative to the ground-truth transition, discouraging overly long or short descriptions that tend to introduce unsupported or noisy UI changes. We optimize the textual transition model using a relative preference objective based on Group Relative Policy Optimization (GRPO)~\citep{shao2024deepseekmathpushinglimitsmathematical}. Further details regarding the GRPO implementation and reward formulation are in Appendix~\ref{sec:appendix_grpo}.

\vspace{-10pt}
\subsection{World-Model-Guided Test-Time Action Search}
\label{sec:planning}

We evaluate CUWM using world-model-guided test-time action search with a frozen agent policy, following~\citep{luo2025vimo}.
At inference time, the agent proposes a set of candidate actions from the current UI state $s_t$.
CUWM simulates the resulting next UI state for each candidate, and the agent selects a single action based on the predicted outcomes.
The agent policy remains unchanged throughout this process, and the world model is used solely as a simulator.
This \emph{think-then-act} procedure allows decision quality to improve with additional test-time computation, which is particularly important for long-horizon computer-using tasks where errors are costly to reverse.

\section{Experiments}
\subsection{Experimental Setup}
\textbf{Dataset.} We train and evaluate CUWM using data from the GUI-360 dataset~\citep{mu2025gui}, focusing on real desktop software, including Microsoft Word, Excel, and PowerPoint.
Each sample is constructed as a state–action–next-state tuple $(s_t, a_t, s_{t+1})$.
More details about the dataset can be found in Appendix~\ref{appendix:data}.

\textbf{Training.} We implement a two-stage pipeline: LoRA-based SFT for both textual (Qwen2.5-VL) and visual (Qwen-Image-Edit) models, followed by GRPO refinement for the textual model (details in Appendix~\ref{appendix:training_details}).

\textbf{Metrics.} We evaluate CUWM from two perspectives: (i) \emph{world model fidelity}, assessing the quality of predicted transitions independent of agents, and (ii) \emph{agent-level evaluation}, measuring CUWM's impact on agent performance via test-time action search.

\subsection{World Model Fidelity}
In this section, we evaluate whether CUWM accurately captures UI dynamics independent of any downstream agent, focusing on both textual state transitions and visual state realization.

\subsubsection{Textual State Transition Evaluation}
\label{sec:textual_state_transition_evaluation}
We evaluate the textual state transition model 
independently, as it provides the explicit representation of UI dynamics used for visual realization and agent reasoning.
We compare three variants: \textbf{Base}, an untrained Qwen2.5-VL model; \textbf{SFT}, trained with supervised fine-tuning; and \textbf{SFT+RL}, the full CUWM model with additional RL refinement.
We evaluate textual transitions using two complementary metrics.

\textbf{LLM-as-a-Judge Score.}
Predicted transitions are compared against GPT-5 generated ground-truth descriptions derived from $(s_t, s_{t+1})$ using an LLM-as-a-Judge.
The judge evaluates consistency across 
\begin{wraptable}{r}{0.42\linewidth}
\vspace{-10pt}
\centering
\caption{Textual state transition model evaluation: LLM-as-a-Judge Score.}
\vspace{-6pt}
\label{tab:textual_judge}
\setlength{\tabcolsep}{4pt}
\renewcommand{\arraystretch}{1.1}
\begin{tabular}{lccc}
\hline
\textbf{Model} & Base & SFT & SFT+RL \\
\hline
\textbf{Judge Score $\uparrow$} & 0.6027 & 0.6834 & \textbf{0.6883} \\
\hline
\end{tabular}
\vspace{-12pt}
\end{wraptable}
multiple UI aspects (e.g., application state, executed actions, and major UI components), assigning scores of 0, 0.5, or 1 per aspect.
We use GPT-5 as the judge and report the average score across aspects (See more details in Appendix~\ref{sec:appendix_llm_judge}).
As shown in Table~\ref{tab:textual_judge}, scores improve from Base to SFT and further to SFT+RL.

\textbf{Action Consistency Score.} To evaluate whether the generated textual transitions preserve decision-relevant information, we introduce the \textbf{Action Consistency Score(ACS)}. Focusing on a \emph{single-step prediction} setting, this metric measures the functional equivalence between the real UI and the generated text. Specifically, we compute the agreement rate between actions selected by a frozen agent policy when conditioned on two distinct inputs: (i) the ground-truth UI screenshot and (ii) the world model's predicted textual transition(see more details in Appendix~\ref{sec:action_consistency_score}).
We evaluate this metric using two 
\begin{wraptable}{r}{0.43\linewidth}
\vspace{-12pt}
\centering
\caption{Action consistency across agent backbones.}
\vspace{-10pt}
\label{tab:textual_action_consistency}
\small
\setlength{\tabcolsep}{7pt}
\renewcommand{\arraystretch}{1.08}
\begin{tabular}{llc}
\hline
\textbf{Model} & \textbf{Agent} & \textbf{Score} \\
\hline
Base   & GPT-4.1-mini     & 0.4990 \\
Base   & Gemini-2.0-Flash & 0.3860 \\
SFT    & GPT-4.1-mini     & 0.5450 \\
SFT    & Gemini-2.0-Flash & 0.4368 \\
SFT+RL & GPT-4.1-mini     & \textbf{0.5642} \\
SFT+RL & Gemini-2.0-Flash & \textbf{0.4732} \\
\hline
\end{tabular}
\vspace{-10pt}
\end{wraptable}
representative agent backbones, GPT-4.1-mini and Gemini-2.0-Flash, to assess robustness across different multimodal reasoning models.
As shown in Table~\ref{tab:textual_action_consistency}, the SFT+RL variant achieves the highest action consistency across both backbones.
Crucially, this improvement in ACS translates into tangible gains in downstream agent performance (as detailed in Table~\ref{tab:model-comparison}), confirming that RL-driven refinement effectively captures decision-critical UI structures necessary for accurate planning.

\subsection{Visual State Realization Evaluation}

The visual state realization model is a critical component of CUWM, as it translates predicted textual state transitions into concrete UI screenshots that can be directly perceived by agents. We evaluate this model using two complementary criteria: image-based quality metrics and text perception score, which together assess visual fidelity and the correctness of rendered UI text.
We study how different sources of textual state transitions affect visual realization quality and compare our approach against the off-the-shelf Qwen-Image-Edit-2509 model. 
Specifically, we consider four evaluation settings: 
(1) \textbf{Action-Only + Qwen-Edit}: directly conditions Qwen-Image-Edit-2509 on the UI screenshot and action.

(2) \textbf{Base-Text + Qwen-Edit}: conditions Qwen-Image-Edit-2509 on textual state transitions generated by the Base textual state transition model.
(3) \textbf{SFT-Text + Qwen-Edit}: conditions Qwen-Image-Edit-2509 on textual state transitions generated by the SFT textual state transition model.
(4) \textbf{SFT-Text + Finetuned-Visual (CUWM)}: conditions a finetuned visual state realization model on SFT textual transitions and corresponds to the full CUWM setting.
\textbf{Image-Based Metrics} To evaluate the similarity between the generated and ground-truth next-state UI screenshots, 
we adopt standard image quality metrics, 
\begin{wraptable}{r}{0.52\linewidth}
\vspace{-8pt}
\centering
\caption{Visual state realization: image-based metrics.}
\vspace{-8pt}
\label{tab:image_metrics}
\footnotesize
\setlength{\tabcolsep}{5pt}
\renewcommand{\arraystretch}{1.08}
\begin{tabular}{lcccc}
\hline
\textbf{Method} 
& \textbf{PSNR} 
& \textbf{SSIM} 
& \textbf{LPIPS} 
& \textbf{FID} \\
\hline
Action-Only + Edit
& 11.09 & 0.49 & 0.48 & 136.14 \\

Base-Text + Edit
& 12.45 & 0.53 & 0.39 & 32.21 \\

SFT-Text + Edit
& 12.86 & 0.54 & 0.39 & 34.59 \\

\textbf{CUWM}
& \textbf{14.91} 
& \textbf{0.67} 
& \textbf{0.21} 
& \textbf{20.48} \\
\hline
\end{tabular}
\vspace{-12pt}
\end{wraptable}
including PSNR~\citep{hore2010image}, SSIM~\citep{wang2004image}, LPIPS~\citep{zhang2018unreasonableeffectivenessdeepfeatures}, and FID~\citep{heusel2018ganstrainedtimescaleupdate} (details in Appendix~\ref{sec:appendix_metrics}). 
As shown in Table~\ref{tab:image_metrics}, incorporating textual state transitions significantly improves visual fidelity compared to directly conditioning on the previous screenshot and action. Performance is further enhanced by fine-tuning the textual transition model, while jointly fine-tuning both textual and visual components (CUWM) achieves the best results across all metrics, demonstrating superior accuracy and perceptual faithfulness in next-state UI generation.

\textbf{Text Perception Score} Text perception is critical in CUWM, as UI applications rely heavily on textual content to convey semantics.
We therefore evaluate the readability and semantic consistency of rendered UI text across state transitions using an automated vision-based parser~\citep{lu2024omniparserpurevisionbased}(details in Appendix~\ref{sec:appendix_omniparser_score}).

\begin{wraptable}{r}{0.55\linewidth}
\vspace{-8pt}
\centering
\caption{Visual state realization: text perception accuracy.}
\vspace{-8pt}
\label{tab:text_rendering}
\footnotesize
\setlength{\tabcolsep}{7pt}
\renewcommand{\arraystretch}{1.08}
\begin{tabular}{lcccc}
\hline
& \multicolumn{4}{c}{\textbf{Text Perception $\uparrow$}} \\
\textbf{Method} 
& \textbf{Word} 
& \textbf{Excel} 
& \textbf{PPT} 
& \textbf{Overall} \\
\hline
Action-Only + Edit
& 0.314 & 0.269 & 0.339 & 0.307 \\

Base-Text + Edit
& 0.621 & 0.614 & 0.542 & 0.597 \\

SFT-Text + Edit
& 0.591 & 0.655 & 0.455 & 0.574 \\

\textbf{CUWM}
& \textbf{0.742}
& \textbf{0.707}
& \textbf{0.689}
& \textbf{0.716} \\
\hline
\end{tabular}
\vspace{-12pt}
\end{wraptable}
Table~\ref{tab:text_rendering} reports results across Word, Excel, and PowerPoint. CUWM achieves the best performance across all applications, indicating that jointly fine-tuning the textual state predictor and the visual renderer substantially improves text preservation during UI transitions. As shown in Figure~\ref{fig:text_rendering_score_curve}, the Text Perception Score generally increases over training epochs.

\subsection{World-Model-Guided Test-Time Action Search}

In this section, we evaluate the impact of the proposed CUWM on agent performance.
We follow the agent design and test-time planning procedure described in Section~\ref{sec:planning}, and use it as our evaluation protocol.

We evaluate CUWM through controlled comparisons across different world model configurations and agent backbones.
Specifically, we consider four agent backbones: Qwen3-VL-8B~\citep{yang2025qwen3technicalreport}, GPT-4.1-mini~\citep{openai2024gpt4technicalreport}, GPT-4o~\citep{openai2024gpt4technicalreport}, and Gemini-2.0-Flash~\citep{geminiteam2025geminifamilyhighlycapable}.

\begin{table*}[!htbp]
\centering
\small
\caption{Agent task scores across different visual state realization models for different agents.}
\vspace{-6pt}
\label{tab:agent-eval}
\setlength{\tabcolsep}{5pt}
\renewcommand{\arraystretch}{1.12}
\begin{tabular}{lccccccccc}
\hline
\multirow{2}{*}{\textbf{Agent}} 
& \multirow{2}{*}{\textbf{None}}
& \multirow{2}{*}{\textbf{Text}}
& \multicolumn{2}{c}{\textbf{Qwen-Edit}}
& \multicolumn{2}{c}{\textbf{CUWM (ours)}} 
& \multicolumn{2}{c}{\textbf{GPT-Image}} \\
\cline{4-9}
& & & Image & Image+Text & Image & Image+Text & Image & Image+Text \\
\hline
Qwen3-VL-8B      & 0.3895 & 0.4102 & 0.4051 & 0.4120 & \textbf{0.4189} & 0.4137 & 0.4137 & 0.4080 \\
GPT-4.1-mini     & 0.4361 & 0.4279 & 0.4196 & 0.4286 & \textbf{0.4418} & 0.4089 & 0.4189 & 0.4127 \\
GPT-4o           & 0.4558 & 0.4625 & 0.4506 & 0.4668 & \textbf{0.4720} & 0.4624 & 0.4514 & 0.4587 \\
Gemini-2.0-Flash & 0.3923 & 0.4008 & 0.4053 & 0.3728 & \textbf{0.4073} & 0.3649 & 0.4004 & 0.3821 \\
\hline
\end{tabular}
\end{table*}

For each backbone, we examine variants without a world model (None), with the Textual State Transition Model only, with the Visual State Realization Model only, as well as their combinations (CUWM) under different integration strategies.
In addition, we compare CUWM with two representative image-generation-based world model baselines, Qwen-Image-Edit-2509~\citep{wu2025qwenimagetechnicalreport} and GPT-Image-1.5~\citep{openai_gpt_image_1_5}.
Table~\ref{tab:agent-eval} reports agent task completion rates under these settings. 
CUWM with image-only input improves performance across all agent backbones, with gains of 4\% for GPT-4o and 8\% for Qwen3-VL-8B. 
These results demonstrate the effectiveness of our proposed world model in enhancing GUI agent decision-making. 
Compared with existing world models, our image-based methods consistently outperform both text-based world models and image-generation baselines across all agents. Further details on the agent evaluation methodology and calculations are provided in Appendix~\ref{appendix:appendix_agent_eval}.

Contrary to expectations, combining text and image predictions degraded agent performance across most configurations. We hypothesize two potential explanations: (1) cross-modal conflict, where textual descriptions contradict visually salient elements, forcing agents to decide between inconsistent signals without a learned resolution strategy, and (2) noise accumulation, where independent prediction errors in each modality compound rather than complement each other when provided together. These findings highlight limitations in current VLMs' capacity for integrated, multimodal reasoning.

\begin{wrapfigure}{r}{0.55\linewidth}
\vspace{-3pt}
\centering
\includegraphics[width=\linewidth]{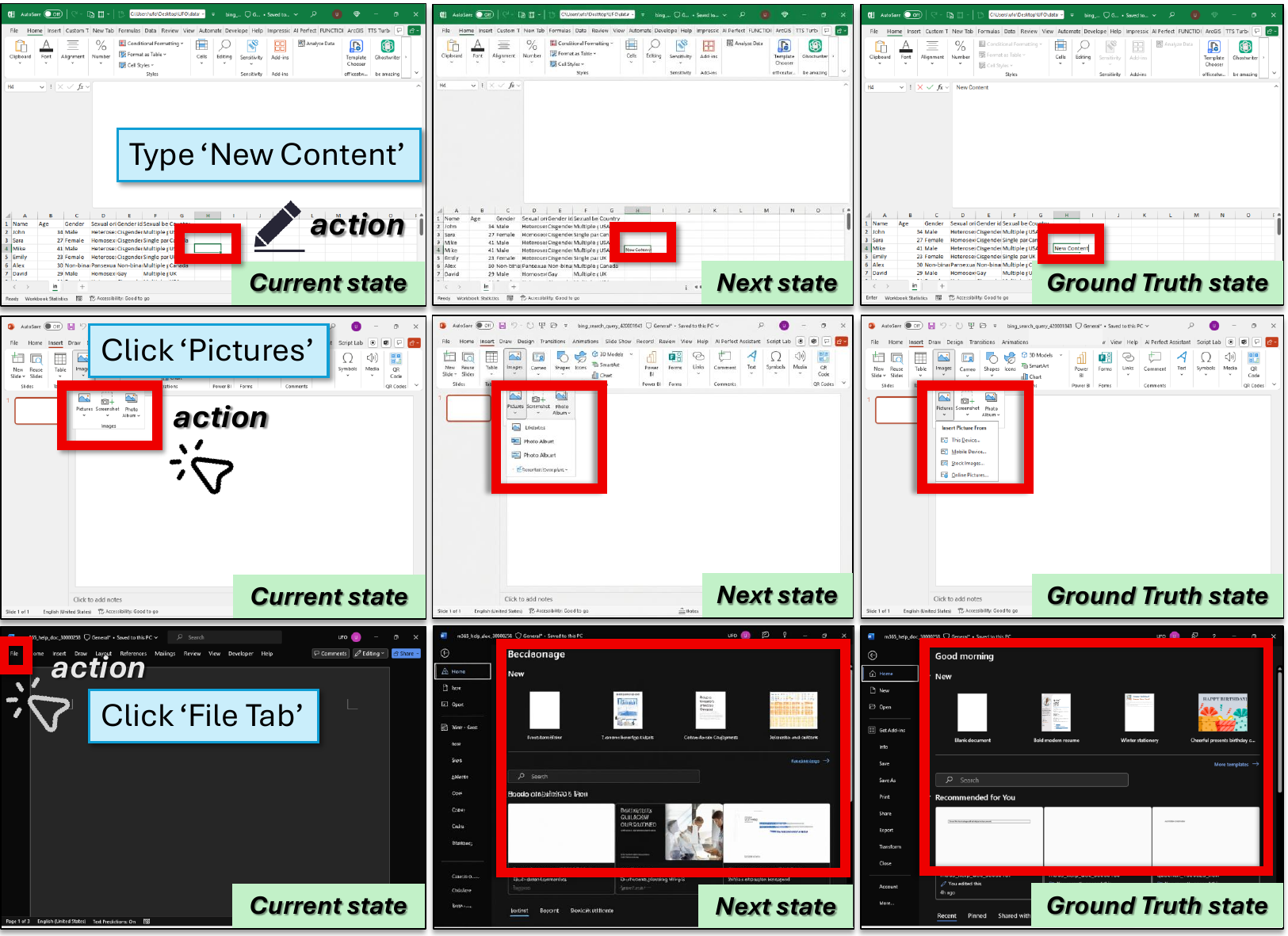}
\caption{Qualitative comparison of CUWM predictions and ground truth under representative UI actions, showing close alignment in both layout and panel states.} 
\label{fig:pred_gt}
\vspace{-10pt}
\end{wrapfigure}

\subsection{Case Study}

\begin{figure}
\vspace{-12pt}
\centering
\includegraphics[width=\linewidth]{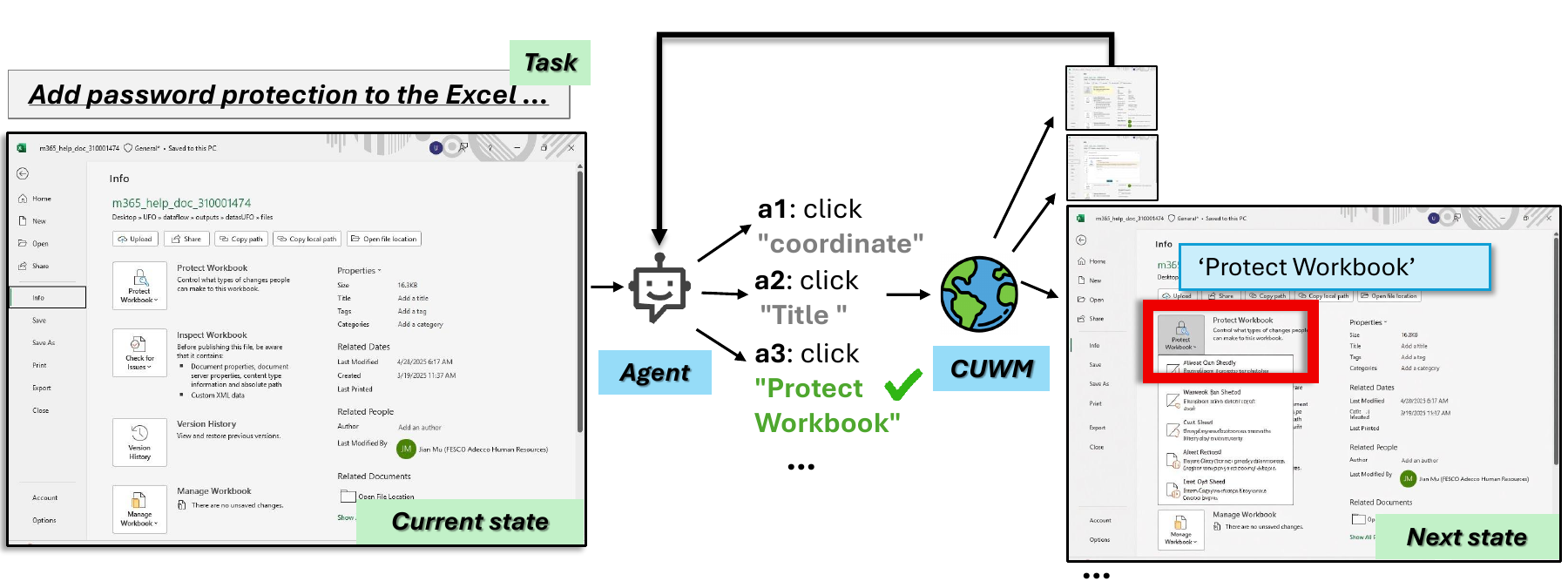}
\caption{\textbf{World-model-guided action selection.}
Given the current Excel UI state and candidate actions, CUWM correctly simulates the respective next states for each action, guiding the agent to select “Protect Workbook” based on goal alignment.}
\label{fig:agent_selection_case_1}
\vspace{-10pt}
\end{figure}

\textbf{Case 1: Capturing Structurally Salient UI Changes.} Figure~\ref{fig:pred_gt} shows cases where the predicted UI state closely matches the ground-truth next state and faithfully reflects the underlying changes. CUWM correctly captures action-induced updates such as text entry, tab switches (e.g., \emph{Pictures}), and opening the \emph{File} view. By accurately modeling these structural transitions and rendering realistic next screens, CUWM enables the agent to anticipate the updated interaction context.
\textbf{Case 2: World-Model-Guided Test-Time Action Search.} Figure~\ref{fig:agent_selection_case_1} illustrates how CUWM supports action selection by \emph{simulating} the outcomes of multiple candidates before execution. Given the task “Add password protection to the Excel workbook”, the agent proposes several candidate actions (e.g., clicking “coordinate”, “Title”, or “Protect Workbook”). Then CUWM correctly simulates the subsequent states for each candidate respectively, providing accurate visual evidence of their distinct outcomes. Guided by these predictions, the agent identifies “Protect Workbook” as the optimal action consistent with the goal, effectively precluding live trial-and-error.

\subsection{Insights: How World Models help GUI agents}
A dominant failure mode of VLM-based agents is their inability to anticipate the consequences of actions \citep{shi2026gui}, which often leads to ineffective or repetitive interactions. CUWM directly addresses this by enabling agents to reason over predicted post-action UI states before execution. We find that world models are especially valuable for structural UI transitions, opening modals, expanding dropdowns, or activating side panes, where changes are visually localized but fundamentally alter the interaction context and valid subsequent actions. By explicitly predicting such transitions, CUWM provides actionable signals about interface changes that VLMs frequently fail to infer from the current state alone.
A notable insight is that agent performance correlates more strongly with access to high-level structural information (e.g., ``a dropdown menu appeared") than with pixel-level fidelity (e.g., ``a precise icon rendered in the dropdown"). This explains why CUWM-based agents outperform those using GPT-Image-1.5 (Table~\ref{tab:agent-eval}) despite lower visual fidelity metrics (Table~\ref{tab:image_metrics}). Finally, world-model-based simulation mitigates a common planning failure: action loops where agents repeatedly select actions that return the interface to its current state. By previewing outcomes, agents can distinguish actions that advance toward the goal from those resulting in stagnation.

\section{Conclusion}

We introduced the Computer-Using World Model (CUWM), a two-stage world model for desktop productivity software that factorizes UI dynamics into textual transition prediction and visual state realization.
This design captures structurally salient UI changes while remaining compatible with pixel-level agent interaction, and can be trained from offline UI transitions with lightweight refinement.
Across a range of Microsoft Office tasks, CUWM serves as an effective test-time simulator for computer-using agents, enabling world-model-guided action search that improves decision quality and execution robustness without modifying agent policies.
Importantly, these gains hold even in deterministic software environments, highlighting the value of test-time simulation for reliable computer use.
Several promising directions could be explored in future work.
One potential avenue is to incorporate reinforcement learning on top of DiT fine-tuning to further align the world model with downstream decision-making objectives.
Another direction is to design reward functions that more directly reflect the usefulness of the world model for agent performance.
Finally, improving the joint training of textual and visual components may help better preserve decision-relevant information during state transitions.

\bibliography{iclr2026_conference}
\bibliographystyle{iclr2026_conference}

\appendix
\section{Appendix}

\subsection{Dataset Summarization}
\label{appendix:data}

Our dataset is constructed based on GUI-360~\citep{mu2025gui}. From this dataset, we sampled and processed task trajectories across three widely used Office applications: Word, Excel, and PowerPoint. GUI-360 provides frame-by-frame screenshots for each trajectory together with rich additional metadata, including the corresponding action commands at each step and task completion signals.

During data preprocessing, we paired each current UI screenshot with the executed action and the subsequent UI screenshot, forming a tuple $(s_t, a_t, s_{t+1})$ as one sample. We select successful trajectories from the original dataset and uniformly sample from those with continuous trajectories and complete, standardized action annotations. In total, we collected 2,876 samples for training and 339 samples for evaluation, serving as an initial dataset that can be readily scaled up in future work. To facilitate efficient and stable training, we standardized image resolutions and removed samples where the pre- and post-action images remained unchanged, the action was invalid, or the data contained excessive noise.
We organize the dataset into training, validation, and test splits, and report the number of samples for each split in Table~\ref{tab:data_split}.

\begin{table}[!htbp]
    \centering
    \caption{Number of samples in the CUWM dataset for each data split and application.}
    \label{tab:data_split}
    \renewcommand{\arraystretch}{1.12}
    \begin{tabular}{lccc}
        \toprule
        \textbf{Data Split} & \textbf{Word} & \textbf{Excel} & \textbf{PowerPoint} \\
        \midrule
        Training    & 797 & 997 & 1082 \\
        Validation  & 40  & 31  & 27   \\
        Test        & 119 & 96  & 124  \\
        \midrule
        Total       & 956 & 1124 & 1233 \\
        \bottomrule
    \end{tabular}
\end{table}

\subsection{Training Details}
\label{appendix:training_details}

\subsubsection{Overview of the training pipeline}
\label{sec:appendix_pipeline}
Our Computer-Using World Model (CUWM) factorizes UI dynamics into two stages: 
(i) a \emph{textual state transition model} that predicts a concise transition description $\Delta_t$ from the current UI screenshot and action $(s_t, a_t)$, and 
(ii) a \emph{visual state realization model} that synthesizes the next UI screenshot $\hat{s}_{t+1}$ conditioned on the current UI and the predicted transition $(s_t, \Delta_t)$.
We train these two stages in a staged manner:
\textbf{(1) supervised initialization (SFT)} for both Stage~1 and Stage~2, subsequently applying \textbf{(2) reinforcement learning refinement (GRPO)} to Stage~1 only.

\begin{table*}[!htbp]
    \vspace{-6pt}
    \centering
    \small
    \caption{Stage 1 (Qwen2.5-VL) SFT hyperparameters.}
    \label{tab:sft_hparams_qwen25vl_lora}
    \renewcommand{\arraystretch}{1.12}
    \begin{tabular}{ll}
    \toprule
    \textbf{Category} & \textbf{Setting} \\
    \midrule
    Model & Qwen2.5-VL-7B-Instruct \\
    Train type & LoRA \\
    Target modules & all-linear \\
    LoRA rank & 32 \\
    LoRA alpha & 32 \\
    Precision & bfloat16 \\
    Learning rate & 1e-4 \\
    Warmup ratio & 0.05 \\
    Weight decay & 0 \\
    Batch size & 4 \\
    Grad accumulation & 4 \\
    \bottomrule
    \end{tabular}
    \vspace{-8pt}
\end{table*}

\subsubsection{Stage 1: Qwen2.5-VL supervised training (textual transitions)}
\label{sec:appendix_s1_sft}
In this stage, we use Qwen2.5-VL as a vision-language model to map $(s_t, a_t)$ to a textual transition description $\Delta_t$.
The model receives the current UI screenshot $s_t$ and a natural-language action $a_t$ as input, and is trained to generate $\Delta_t^{\mathrm{GT}}$. We optimize a standard autoregressive cross-entropy loss on the target transition text:
$\mathcal{L}_{\mathrm{SFT}} = -\log p(\Delta_t^{\mathrm{GT}} \mid s_t, a_t)$.
We fine-tune the model using LoRA,
all Stage~1 hyperparameters are summarized in Table~\ref{tab:sft_hparams_qwen25vl_lora}.

\begin{table}[!htbp]
    \centering
    \small
    \renewcommand{\arraystretch}{1.12}
    \setlength{\tabcolsep}{6pt}
    \caption{Stage 2 (Qwen-Image-Edit) SFT hyperparameters.}
    \label{tab:sft_hparams_qwen_image_edit_lora}
    \begin{tabular}{l p{0.6\linewidth}}
    \toprule
    \textbf{Category} & \textbf{Setting} \\
    \midrule
    Model & Qwen-Image-Edit-2509 \\
    Training type & LoRA \\
    LoRA base & DiT \\
    Target modules &
    \texttt{to\_q, to\_k, to\_v, add\_q\_proj, add\_k\_proj,} \\
    & \texttt{add\_v\_proj, to\_out.0, to\_add\_out, img\_mlp.net.2,} \\
    & \texttt{img\_mod.1, txt\_mlp.net.2, txt\_mod.1} \\
    LoRA rank & 32 \\
    LoRA alpha & 32 \\
    Precision & 16-mixed \\
    Learning rate & 1e-4 \\
    Warmup ratio & 0 \\
    Weight decay & 0.01 \\
    Batch size & 1 \\
    Grad accumulation & 1 \\
    Max pixels & 1{,}048{,}576 \\
    \bottomrule
    \end{tabular}
\end{table}

\begin{table}[!htbp]
    \centering
    \small
    \caption{Stage 1 GRPO hyperparameters.}
    \label{tab:hparams_grpo}
    \renewcommand{\arraystretch}{1.12}
    \begin{tabular}{ll}
    \toprule
    \textbf{Category} & \textbf{Setting} \\
    \midrule
    Model & SFT-Qwen2.5-VL \\
    Algorithm & GRPO \\
    Train type & LoRA \\
    Target modules & all-linear \\
    Exclude modules & .*visual.* \\
    KL loss & enabled (coef=0.01) \\
    LoRA rank & 64 \\
    LoRA alpha & 32 \\
    Precision & bfloat16 \\
    Learning rate & 3e-6 \\
    Warmup ratio & 0 \\
    Weight decay & 0.01 \\
    Train batch size & 32 \\
    Rollout $n$ & 5 \\
    ppo mini batch size & 32 \\
    ppo micro batch size per gpu & 8 \\
    \midrule
    Length Penalty Weight ($\beta$) & 1 \\
    Min Length Ratio ($r_{\text{low}}$) & 0.75 \\
    Max Length Ratio ($r_{\text{up}}$) & 1.25 \\
    Max Penalty Score ($m$) & 1.0 \\
    \bottomrule
    \end{tabular}
\end{table}

\subsubsection{Stage 2: Qwen-Image-Edit supervised training (visual realization)}
\label{sec:appendix_s2_sft}
In this stage, we use Qwen-Image-Edit as a conditional image editing model to generate the next UI screenshot.
The model is conditioned on the current UI screenshot and the predicted textual transition:
$\hat{s}_{t+1} = f_{\mathrm{image}}(s_t, \Delta_t)$.
We fine-tune the model with a pixel-wise reconstruction objective using the mean squared error (MSE) loss:
$\mathcal{L}_{\mathrm{EDIT}} = \| \hat{s}_{t+1} - s_{t+1} \|_2^2$,
where $s_{t+1}$ denotes the ground-truth next-state screenshot.
We apply LoRA fine-tuning only to the DiT backbone of Qwen-Image-Edit, 
all Stage~2 hyperparameters are summarized in Table~\ref{tab:sft_hparams_qwen_image_edit_lora}.

\subsubsection{Stage 1 RL Refinement: Qwen2.5-VL + GRPO}
\label{sec:appendix_grpo}

Although SFT provides a faithful initialization, it does not guarantee that the generated textual transition $\Delta_t$ consistently captures the most decision-critical UI structures.
To address this, we further refine the Stage~1 model using Group Relative Policy Optimization (GRPO), treating the VLM as a policy $\pi_\theta$ that generates $\Delta_t$ conditioned on $(s_t, a_t)$.

During training, for each input tuple $(s_t, a_t)$, we sample a group of $K=5$ candidate descriptions $\{\Delta_t^{(k)}\}_{k=1}^K$ from the current policy using a temperature of \textbf{1.0} and top-$p$ of \textbf{1.0}.
GRPO optimizes the relative preference among these $K$ samples without requiring a separate value network (critic), which ensures stability for text generation tasks involving structured metrics.

\paragraph{Reward Formulation.}
The optimization objective is driven by a composite scalar reward function. For a sampled description $\Delta_t$, the total reward is defined as:
\begin{equation}
R(s_t, a_t, \Delta_t) = R_{\mathrm{judge}}(\Delta_t, \Delta_t^{\mathrm{GT}}) - \beta \cdot R_{\mathrm{len}}(\Delta_t, \Delta_t^{\mathrm{GT}}),
\label{eq:grpo_reward}
\end{equation}
where:
\begin{itemize}
    \item $R_{\mathrm{judge}}$ is the semantic consistency score assigned by the LLM-as-a-Judge (as detailed in Section~\ref{sec:appendix_llm_judge});
    \item $R_{\mathrm{len}}$ is a soft length penalty designed to discourage verbosity;
    \item $\beta$ is a hyperparameter controlling the strength of the length regularization.
\end{itemize}

\paragraph{Soft Length Penalty ($R_{\mathrm{len}}$).}
To prevent the model from generating hallucinations through excessive verbosity or missing details due to brevity, we impose a dynamic length penalty.
Let $L_{\text{pred}}$ denote the token count of the predicted description and $L_{\text{gt}}$ denote the token count of the ground truth.
We define a valid length interval $[l_{\text{min}}, l_{\text{max}}]$ relative to the ground truth:
\begin{equation}
l_{\text{min}} = \max\!\left(1, \lfloor r_{\text{low}} \cdot L_{\text{gt}} \rfloor \right), \quad
l_{\text{max}} = \max\!\left(l_{\text{min}} + 1, \lfloor r_{\text{up}} \cdot L_{\text{gt}} \rfloor \right),
\end{equation}
where $r_{\text{low}}$ and $r_{\text{up}}$ are scaling factors defining the acceptable range. The penalty term $R_{\mathrm{len}}$ increases linearly with the deviation from this interval, capped by a maximum penalty $m$:
\begin{equation}
R_{\mathrm{len}} =
\begin{cases}
0, & l_{\text{min}} \le L_{\text{pred}} \le l_{\text{max}}, \\
m \cdot \min\!\left(1, \frac{l_{\text{min}} - L_{\text{pred}}}{l_{\text{min}}} \right), & L_{\text{pred}} < l_{\text{min}}, \\
m \cdot \min\!\left(1, \frac{L_{\text{pred}} - l_{\text{max}}}{l_{\text{max}}} \right), & L_{\text{pred}} > l_{\text{max}}.
\end{cases}
\label{eq:length_penalty}
\end{equation}
This formulation encourages concise yet structurally complete transition descriptions. Detailed hyperparameters for GRPO training are summarized in Table~\ref{tab:hparams_grpo}.

\begin{figure*}[!htbp]
  \centering
  \includegraphics[width=0.7\linewidth]{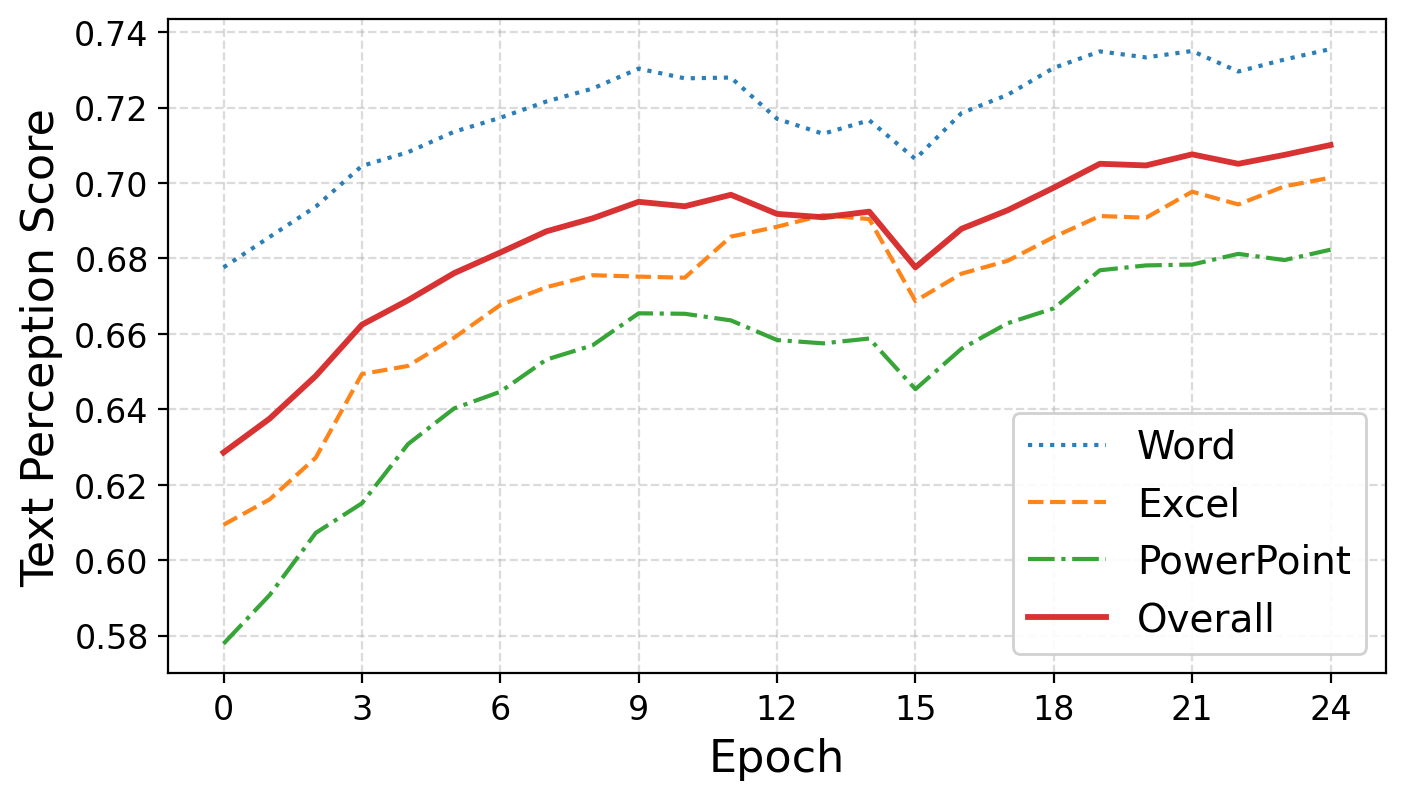}
  \caption{Training curve over epochs for Text Perception Score ($\uparrow$).}
    \label{fig:text_rendering_score_curve}
\end{figure*}

\subsubsection{Training Curves and Analysis}
\label{sec:appendix_curves}

\begin{figure*}[!htbp]
    \centering
    \captionsetup[subfigure]{font=small}
    \begin{subfigure}[t]{0.48\textwidth}
        \centering
        \includegraphics[width=\linewidth]{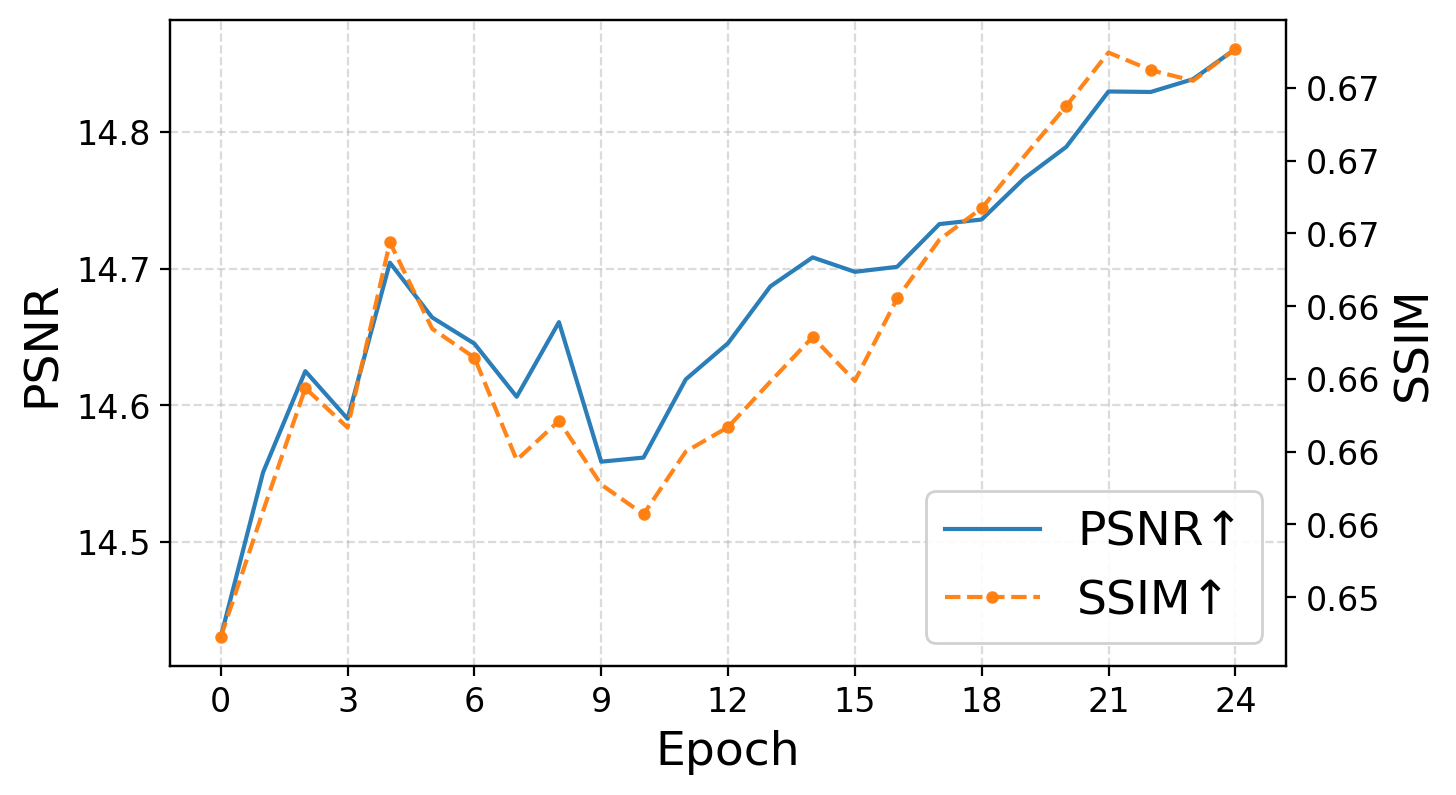}
        \label{fig:psnr_ssim_vs_epoch_curve}
    \end{subfigure}
    \hfill
    \begin{subfigure}[t]{0.48\textwidth}
        \centering
        \includegraphics[width=\linewidth]{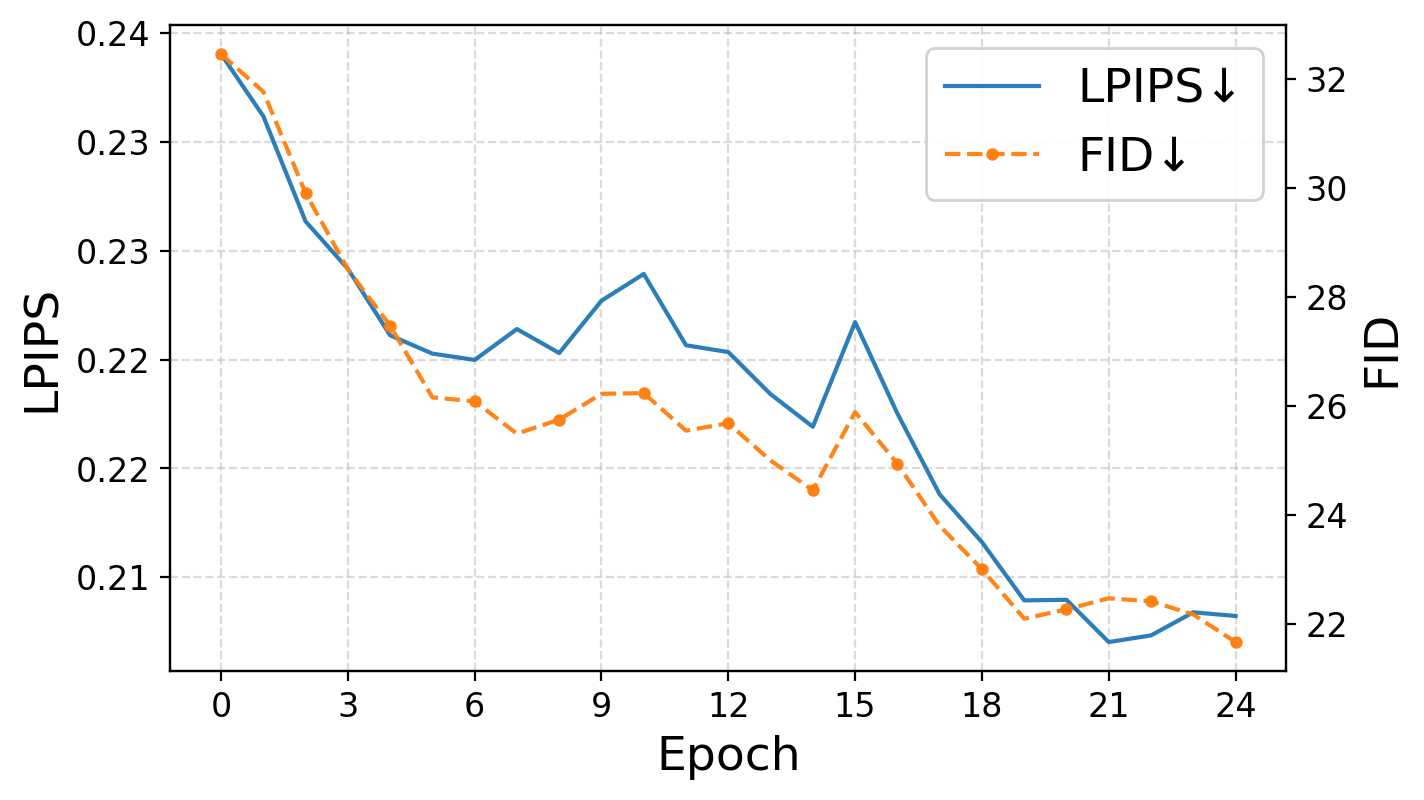}
        \label{fig:lpips_fid_vs_epoch_curve}
    \end{subfigure}
    \caption{Training curves over epochs for image-level fidelity metrics.}
    \label{fig:image_metrics}
\end{figure*}

Given the inherent stochasticity of diffusion model training, simple loss functions (e.g., MSE) are often insufficient proxies for generation quality. Therefore, we monitor the training progress using the comprehensive set of image-based metrics and the text perception score defined in Section~\ref{sec:evaluation}.
Figure~\ref{fig:text_rendering_score_curve} and Figure~\ref{fig:image_metrics} illustrate the training dynamics across epochs.

As shown in Figure~\ref{fig:image_metrics}, we observe consistent improvements in visual fidelity: \textbf{PSNR} and \textbf{SSIM} exhibit an upward trend, while \textbf{LPIPS} and \textbf{FID} steadily decrease. This trajectory indicates that the model progressively enhances pixel-level accuracy, structural alignment, and perceptual realism.
Aligned with these visual gains, Figure~\ref{fig:text_rendering_score_curve} demonstrates that the \textbf{Text Perception Score} increases steadily across all applications. This confirms that improvements in general image quality effectively translate into more accurate rendering of legible UI text and application-specific patterns.
Furthermore, a distinct performance gap emerges across domains: \textbf{Word} consistently achieves the highest scores, followed by \textbf{Excel}, with \textbf{PowerPoint} proving the most challenging. This stratification likely reflects the varying degrees of UI layout complexity and text density inherent to each application.

\subsection{EVALUATION}
\label{sec:evaluation}
\subsubsection{LLM-as-a-Judge Score}
\label{sec:appendix_llm_judge}

Standard pixel-level similarity metrics often fail to capture the semantic nuances of \emph{what actually changed} in the UI following an action (e.g., updates to the active tab, auxiliary panes, or document content).
To address this, we employ an \emph{LLM-as-a-Judge} protocol designed to quantify the semantic consistency between the model's predicted textual description and a reference description.

For each transition, we compare the model-predicted description $\hat{y}$ against a ground-truth description $y$.
We utilize GPT-5 to generate $y$ from the paired screenshots $(s_t, s_{t+1})$, establishing it as the single source of truth.
Given the pair $(\hat{y}, y)$, the judge (GPT-5) is prompted to output a structured JSON containing discrete scores and rationales for specific UI aspects. The complete evaluation prompt is provided in Appendix~\ref{sec:llm_judge_prompt}.

\begin{table}[!htbp]
\vspace{-4pt}
\centering
\caption{LLM-as-a-Judge evaluation aspects and their weights.}
\label{tab:judge_aspects_weights}
\setlength{\tabcolsep}{6pt}
\renewcommand{\arraystretch}{1.15}
\begin{tabular}{llc}
\hline
\textbf{Key} & \textbf{UI aspect} & \textbf{Weight $w_a$} \\
\hline
\texttt{app\_name} & Application name & 0.8 \\
\texttt{user\_action} & Executed user action & 1.4 \\
\texttt{title\_bar} & Title bar state (e.g., document name) & 1.0 \\
\texttt{ribbon} & Ribbon / toolbar state (e.g., active tab) & 1.1 \\
\texttt{main\_editing\_area} & Main editing area / canvas (core content changes) & 1.5 \\
\texttt{sidebar\_pane} & Sidebar / pane state (e.g., open/closed, content) & 0.8 \\
\texttt{navigation\_area} & Navigation area (e.g., thumbnails/outline) & 0.6 \\
\texttt{status\_bar} & Status bar (e.g., page, zoom, mode) & 0.8 \\
\hline
\end{tabular}
\vspace{-8pt}
\end{table}

\paragraph{Per-aspect discrete scoring.}
Let $\mathcal{A}$ denote the set of UI aspects (e.g., \texttt{user\_action}, \texttt{main\_editing\_area}) as listed in Table~\ref{tab:judge_aspects_weights}, and let $a \in \mathcal{A}$ represent a specific aspect.
For a generated description $\hat{y}$ and the ground truth $y$, the aspect-specific score $s_a(\hat{y},y)$ is determined as:

\begin{equation}
s_a(\hat{y}, y) =
\begin{cases}
1, & \text{if } \hat{y} \text{ aligns with } y \text{ on key elements with high fidelity;} \\
0.5, & \text{if } \hat{y} \text{ is partially correct but has notable omissions;} \\
0, & \text{if } \hat{y} \text{ contradicts } y \text{ or introduces hallucinations.}
\end{cases}
\label{eq:discrete_scoring}
\end{equation}

The judge is explicitly instructed to prioritize \emph{content fidelity} over stylistic variations. Omissions of sub-areas not mentioned in the reference are treated neutrally, whereas asserting incorrect changes results in a penalty (score 0).

\paragraph{Aspect weights and final score.}
To reflect the hierarchical importance of different UI components in downstream reasoning, we compute a weighted average of the aspect scores:
\begin{equation}
\mathrm{JudgeScore}(\hat{y},y)
=
\frac{\sum_{a\in\mathcal{A}} w_a\, s_a(\hat{y},y)}
{\sum_{a\in\mathcal{A}} w_a}.
\label{eq:judge_score_weighted}
\end{equation}
As detailed in Table~\ref{tab:judge_aspects_weights}, this weighting scheme places higher emphasis on the \texttt{main\_editing\_area} and \texttt{user\_action}, as these are critical for task progression, while still accounting for the global UI context (e.g., ribbon and title bar).We apply this metric to evaluate performance across different training stages (Base, SFT, SFT+RL), with results summarized in Table~\ref{tab:textual_judge}.

\subsubsection{Action Consistency Score}
\label{sec:action_consistency_score}

\begin{figure}[htbp]
\centering
\begin{lstlisting}[style=whitebg]
  {
    "function": "select_text",
    "args": { "text": "artificial intelligence (AI)" },
    "status": "CONTINUE"
  },
  {
    "function": "click",
    "args": {
      "coordinate": null,
      "button": "left",
      "control_info": {
        "control_type": "Button",
        "control_text": "Text Highlight Color Yellow"
      }
    },
    "status": "FINISH"
  }
]
\end{lstlisting}
\caption{\textbf{Example action format.} Two candidate actions illustrating the structured JSON specification used by agents. The first action selects text, while the second clicks a UI control identified by its accessibility label.}
\label{fig:action_format}
\end{figure}

A faithful textual state transition model must preserve \emph{decision-relevant} information from the user interface. To quantify this, we introduce the \textbf{Action Consistency Score (ACS)}, which measures the functional equivalence between the generated textual state and the ground-truth visual state.
The core intuition is that an agent acting solely on the world model's textual prediction should arrive at the same decision as an agent observing the actual screenshot.

\textbf{Evaluation Protocol.}
Formally, each evaluation instance consists of a tuple $(I, D_{\mathrm{gt}}, A, T)$, where $I$ is the ground-truth current screenshot, $D_{\mathrm{gt}}$ is the ground-truth textual description derived from $I$, $A$ represents accessibility metadata (e.g., control labels, application name), and $T$ is the user instruction.
Given a world model, we generate a predicted textual transition $D_{\mathrm{wm}}$ for the next state.
We employ a \emph{frozen} agent policy $\pi$ to predict the next action under two distinct conditions:
\begin{align}
a_{\mathrm{gt}} &= \pi\big(I,\; D_{\mathrm{gt}},\; A,\; T), \label{eq:action_gt} \\
a_{\mathrm{wm}} &= \pi\big(\varnothing,\; D_{\mathrm{wm}},\; A,\; T), \label{eq:action_wm}
\end{align}
Here, $a_{\mathrm{gt}}$ serves as the \emph{oracle action} derived from the full visual context $I$, whereas $a_{\mathrm{wm}}$ is the \emph{predicted action} derived exclusively from the world model's textual output $D_{\mathrm{wm}}$ without visual access.

\textbf{Scoring Formulation.}
To compute consistency, we first verify that the agent's output adheres to a structured JSON specification. Specifically, each action $a$ is parsed into three key components:
(1) a \texttt{function} field denoting the action type (e.g., \texttt{click}, \texttt{select\_text});
(2) an \texttt{args} dictionary containing parameters such as target coordinates, control identifiers, or text content; and
(3) a \texttt{status} field (\texttt{CONTINUE} or \texttt{FINISH}) indicating task progression. (see Figure~\ref{fig:action_format} for an example))

Based on this structure, we define three atomic matching criteria as detailed in Table~\ref{tab:action_consistency_metric}: \textbf{Function Match}, \textbf{Status Match}, and \textbf{Args Match} (which incorporates spatial tolerance and label matching).
The instance-level consistency score is computed as a weighted sum, prioritizing the correctness of action arguments:
\begin{equation}
s(a_{\mathrm{wm}}, a_{\mathrm{gt}})
= 0.25 \cdot \mathbbm{1}_{\textsc{Func}}
+ 0.25 \cdot \mathbbm{1}_{\textsc{Status}}
+ 0.50 \cdot \mathbbm{1}_{\textsc{Args}}
\label{eq:action_consistency_instance_score}
\end{equation}
If the agent fails to produce a valid action in either condition, we assign $s=0$.
The final \textbf{Action Consistency Score} is the average score over the entire evaluation dataset $\mathcal{D}$:
\begin{equation}
\mathrm{ACS}
= \frac{1}{|\mathcal{D}|}\sum_{(I,D_{\mathrm{gt}},A,T)\in\mathcal{D}}
s\!\left(a_{\mathrm{wm}}, a_{\mathrm{gt}}\right).
\label{eq:action_consistency_score}
\end{equation}

\textbf{Interpretation.}
A high ACS indicates that the textual transition $D_{\mathrm{wm}}$ successfully retains critical UI cues—such as active tabs, control states, or content updates—necessary for accurate decision-making. Conversely, a low ACS implies that the textual abstraction has lost or distorted information vital for the agent's policy, leading to divergent behavior compared to the visual oracle.

\begin{table}[htbp]
\centering
\caption{\textbf{Scoring metrics for action evaluation.} Each metric evaluates a specific aspect of action correctness. The Overall Match criterion implies all component metrics are satisfied.}
\label{tab:action_consistency_metric}
\small
\setlength{\tabcolsep}{6pt}
\renewcommand{\arraystretch}{1.2}
\begin{tabular}{@{}lp{0.72\linewidth}@{}}
\toprule
\textbf{Metric} & \textbf{Description} \\
\midrule
Function Match & Evaluates whether the predicted action type (e.g., \texttt{click}, \texttt{type}, \texttt{drag}, \texttt{select\_text}) exactly matches the ground-truth action type. \\
\addlinespace[4pt]
Status Match & Evaluates whether the predicted task status (\texttt{CONTINUE} or \texttt{FINISH}) matches the ground-truth status, indicating correct anticipation of task completion. \\
\addlinespace[4pt]
Args Match & Evaluates whether the action arguments align with the ground truth. For coordinate-based actions, we allow a spatial tolerance of $\pm 25$ pixels or require the point to fall within the target's bounding box. For label-based actions, exact string matching is required. Additional arguments (e.g., \texttt{button}, \texttt{keys}) must match exactly. \\
\addlinespace[4pt]
Overall Match & A strict criterion requiring \emph{all three} components (Function Match, Status Match, and Args Match) to be satisfied simultaneously. \\
\bottomrule
\end{tabular}
\end{table}

\subsubsection{Visual State Realization Evaluation}
\label{sec:appendix_metrics}

We assess the fidelity of CUWM-generated next-state UI screenshots using four standard image-quality metrics: PSNR, SSIM, LPIPS, and FID.

\noindent\textbf{PSNR} ($\uparrow$).
Peak Signal-to-Noise Ratio measures pixel-wise reconstruction quality based on mean squared error (MSE).
Given a generated image $\hat{\mathbf{I}}$ and the ground-truth image $\mathbf{I}$ of size $H\times W$ (with $C$ channels), the MSE is
\begin{equation}
\mathrm{MSE}(\hat{\mathbf{I}},\mathbf{I})
= \frac{1}{HWC}\sum_{h=1}^{H}\sum_{w=1}^{W}\sum_{c=1}^{C}
\left(\hat{I}_{h,w,c}-I_{h,w,c}\right)^2,
\end{equation}
and PSNR is
\begin{equation}
\mathrm{PSNR}(\hat{\mathbf{I}},\mathbf{I})
= 10 \log_{10}\left(\frac{\mathrm{MAX}^2}{\mathrm{MSE}(\hat{\mathbf{I}},\mathbf{I})}\right),
\end{equation}
where $\mathrm{MAX}$ is the maximum possible pixel value (e.g., $255$ for 8-bit images or $1$ for normalized images).

\noindent\textbf{SSIM} ($\uparrow$).
The Structural Similarity Index evaluates perceptual similarity by comparing luminance, contrast, and structure.
For local windows (patches) from $\hat{\mathbf{I}}$ and $\mathbf{I}$, SSIM is defined as
\begin{equation}
\mathrm{SSIM}(\hat{\mathbf{I}},\mathbf{I})
=
\frac{(2\mu_{\hat{I}}\mu_{I}+c_1)(2\sigma_{\hat{I}I}+c_2)}
{(\mu_{\hat{I}}^2+\mu_{I}^2+c_1)(\sigma_{\hat{I}}^2+\sigma_{I}^2+c_2)},
\end{equation}
where $\mu_{\hat{I}},\mu_{I}$ are local means, $\sigma_{\hat{I}}^2,\sigma_{I}^2$ are local variances,
$\sigma_{\hat{I}I}$ is the local covariance, and $c_1,c_2$ are small constants for numerical stability.
We report SSIM averaged over windows (and channels if applicable).

\noindent\textbf{LPIPS} ($\downarrow$).
Learned Perceptual Image Patch Similarity measures perceptual distance in deep feature space.
Let $\phi^{l}(\cdot)$ denote the feature map from layer $l$ of a pretrained network, normalized channel-wise as $\hat{\phi}^{l}$, and let $\mathbf{w}^{l}$ be learned per-channel weights.
LPIPS is computed as
\begin{equation}
\mathrm{LPIPS}(\hat{\mathbf{I}},\mathbf{I})
=
\sum_{l}\frac{1}{H_l W_l}\sum_{h,w}
\left\|
\mathbf{w}^{l}\odot\left(\hat{\phi}^{l}(\hat{\mathbf{I}})_{h,w}-\hat{\phi}^{l}(\mathbf{I})_{h,w}\right)
\right\|_2^2,
\end{equation}
where $H_l,W_l$ are the spatial dimensions of layer $l$, and $\odot$ denotes element-wise multiplication.

\noindent\textbf{FID} ($\downarrow$).
Fr\'echet Inception Distance measures distribution-level discrepancy between generated and real images in the Inception feature space.
Let $\{\mathbf{x}_i\}$ and $\{\hat{\mathbf{x}}_j\}$ be Inception features for real and generated images, with empirical means and covariances
$(\boldsymbol{\mu}_r,\mathbf{\Sigma}_r)$ and $(\boldsymbol{\mu}_g,\mathbf{\Sigma}_g)$, respectively.
FID is
\begin{equation}
\mathrm{FID}
=
\left\|\boldsymbol{\mu}_r-\boldsymbol{\mu}_g\right\|_2^2
+
\mathrm{Tr}\!\left(
\mathbf{\Sigma}_r+\mathbf{\Sigma}_g
-2\left(\mathbf{\Sigma}_r\mathbf{\Sigma}_g\right)^{\frac{1}{2}}
\right).
\end{equation}

\subsubsection{Text Perception Score}
\label{sec:appendix_omniparser_score}

To evaluate whether a generated next-frame UI screenshot faithfully renders text content (e.g., document text and UI labels), we compute a task-oriented \emph{Text Perception Score} using \textsc{OmniParser} as a screen-text extractor.
Given an image, \textsc{OmniParser} returns a list of parsed elements; we keep only items with \texttt{type = text} and collect their textual \texttt{content} as a multiset of strings.
We apply light normalization (lowercasing, removing duplicates, and filtering short/noisy strings) and denote the resulting text sets from the prediction and the ground truth as
$\mathcal{P}=\{p_i\}_{i=1}^{|\mathcal{P}|}$ and $\mathcal{G}=\{g_j\}_{j=1}^{|\mathcal{G}|}$, respectively.

\paragraph{Embedding similarity.}
We embed each text string with a sentence encoder $\phi(\cdot)$ and use cosine similarity:
\begin{equation}
s_{ij} \;=\; \cos\!\big(\phi(p_i),\,\phi(g_j)\big), \quad p_i \in \mathcal{P},~ g_j \in \mathcal{G}.
\label{eq:omni_cosine}
\end{equation}

\paragraph{Symmetric max-match.}
We measure how well each side can be explained by the other via maximum matching in the embedding space.
For the prediction-to-GT direction:
\begin{equation}
M_{\mathcal{P}\rightarrow\mathcal{G}}
\;=\;
\frac{1}{|\mathcal{P}|}\sum_{i=1}^{|\mathcal{P}|}\max_{j}~ s_{ij},
\label{eq:omni_p2g}
\end{equation}
and for the GT-to-prediction direction:
\begin{equation}
M_{\mathcal{G}\rightarrow\mathcal{P}}
\;=\;
\frac{1}{|\mathcal{G}|}\sum_{j=1}^{|\mathcal{G}|}\max_{i}~ s_{ij}.
\label{eq:omni_g2p}
\end{equation}
The final Text Rendering Score is the symmetric average:
\begin{equation}
\mathrm{TRS}(\mathcal{P},\mathcal{G})
\;=\;
\frac{1}{2}\Big(M_{\mathcal{P}\rightarrow\mathcal{G}} + M_{\mathcal{G}\rightarrow\mathcal{P}}\Big).
\label{eq:omni_trs}
\end{equation}
\textbf{Interpretation.}
$M_{\mathcal{P}\rightarrow\mathcal{G}}$ penalizes hallucinated or irrelevant text in the prediction that cannot find a close match in the ground truth, while
$M_{\mathcal{G}\rightarrow\mathcal{P}}$ penalizes missing or incorrectly rendered text from the ground truth that cannot be recovered from the prediction.
Thus, the symmetric formulation rewards both \emph{precision} (avoiding spurious text) and \emph{recall} (preserving all important text).
If either side is empty, we define
$\mathrm{TRS}(\emptyset,\emptyset)=1$ and $\mathrm{TRS}(\mathcal{P},\emptyset)=\mathrm{TRS}(\emptyset,\mathcal{G})=0$.
We report the dataset-level score by averaging $\mathrm{TRS}$ over all evaluation samples, and optionally report per-application scores (Word/Excel/PowerPoint) by averaging within each subset.

\subsection{Supplementary Agent Evaluation Results}
\label{appendix:appendix_agent_eval}

\subsubsection{Agent Evaluation Protocol}

We evaluate agent performance using a structured protocol designed to isolate the contribution of the world model from the agent policy.
During evaluation, each VLM backbone agent (listed in Table~\ref{tab:agent-eval}) receives the current UI screenshot along with a task instruction.
The agent is prompted to generate five diverse candidate actions matching the structured JSON format illustrated in Figure~\ref{fig:action_format}, using the prompt provided in Appendix~\ref{app:prompts}.
For each candidate action, we pass the task instruction, the action details, and the current screenshot to the CUWM textual transition model, which generates a textual description of the predicted UI changes.
This textual transition is then fed into an image-based world model, either the CUWM visual realization model, the base Qwen-Image-Edit, or GPT-Image-1.5, to render the predicted next-state.
Finally, the agent selects the action whose predicted outcome best aligns with the task goal.

\begin{table*}[!htbp]
\centering
\small
\caption{Agent task completion rates with VLM-specific failures removed to isolate world model contribution. No GT indicates samples where the ground-truth action was absent from the VLM's candidate proposals. \textbf{Bold} indicates best performance per agent}
\vspace{-6pt}
\label{tab:agent-eval_no_gt}
\setlength{\tabcolsep}{5pt}
\renewcommand{\arraystretch}{1}

\resizebox{\textwidth}{!}{%
\begin{tabular}{llccccccccc}
\hline
\multirow{2}{*}{\textbf{Agent}} 
& \multirow{2}{*}{\textbf{No GT}}
& \multirow{2}{*}{\textbf{None}}
& \multirow{2}{*}{\textbf{Text}}
& \multicolumn{2}{c}{\textbf{Qwen-Edit}}
& \multicolumn{2}{c}{\textbf{CUWM (ours)}} 
& \multicolumn{2}{c}{\textbf{GPT-Image}} \\
\cline{5-10}
& & & & Image & Image+Text & Image & Image+Text & Image & Image+Text \\
\hline
Qwen3-VL-8B & 159 & 0.7336 & 0.7725 & 0.7629 & 0.7759 & \textbf{0.7889} & 0.7791 & 0.7791 & 0.7684 \\
GPT-4.1-mini & 115 & 0.6600 & 0.6476 & 0.6350 & 0.6486 & \textbf{0.6686} & 0.6188 & 0.6340 & 0.6246 \\
GPT-4o & 127 & 0.7288 & 0.7396 & 0.7205 & 0.7464 & \textbf{0.7548} & 0.7394 & 0.7218 & 0.7335 \\
Gemini-2.0-Flash & 135 & 0.6519 & 0.6660 & 0.6735 & 0.6195 & \textbf{0.6768} & 0.6064 & 0.6654 & 0.6350 \\
\hline
\end{tabular}%
}
\end{table*}

\subsubsection{Scoring Methodology}

For each test sample, we compare the agent-selected action against the ground-truth action from the GUI-360 dataset using four metrics: \emph{Function Match}, \emph{Status Match}, \emph{Args Match}, and \emph{Overall Match}, summarized in Table~\ref{tab:action_consistency_metric}.
The Overall Match criterion requires that all three preceding metrics agree with the ground truth.
The final agent task score is computed as the proportion of Overall Matches across all 339 evaluation samples as described in Section~\ref{sec:planning}.

\subsubsection{Ground-Truth Action Coverage}
It is important to note that not all candidate action sets generated by the agent include the ground-truth action. We observe that approximately 35\% of evaluation samples fall into this category, though this proportion varies by VLM backbone, each agent independently fails to propose the ground-truth action on different subsets of samples. While this limitation is unrelated to the CUWM framework and reflects inherent agent proposal quality, we retain these samples in the evaluation to provide a faithful measure of end-to-end agent performance. Table~\ref{tab:agent-eval_no_gt} separately reports results excluding these samples, evaluating only cases where the ground-truth action appeared in the candidate set.

\subsubsection{Effectiveness of RL Refinement on Agent task score}
Table \ref{tab:model-comparison} presents the downstream task accuracy averaged over 100 randomly selected samples.
Consistent with the Action Consistency Score (ACS) analysis in Section~\ref{sec:textual_state_transition_evaluation}, the RL-refined CUWM consistently outperforms the SFT baseline across both GPT-4.1-mini and Gemini-2.0-Flash agents.
Specifically, the RL variant achieves higher accuracy on CUWM-generated images ($0.4317$ and $0.4700$, respectively), surpassing the SFT variant.
This validates our hypothesis that the improved semantic preservation achieved via GRPO (indicated by higher ACS) directly translates into higher-fidelity visual realizations that effectively guide agent decision-making.

\begin{table*}[!htbp]
\centering
\caption{\textbf{Agent performance across input modalities.} We report accuracy scores averaged over the last 6 runs with standard deviations where available. \textbf{Bold} indicates best performance per agent. This was tested on 100 randomly selected samples.}

\label{tab:model-comparison}
\setlength{\tabcolsep}{7pt}
\renewcommand{\arraystretch}{1.12}
\resizebox{0.85\textwidth}{!}{%
\begin{tabular}{llccc}
\hline
\textbf{Model} & \textbf{Method} & \textbf{Text} & \textbf{Base Image} & \textbf{CUWM Image} \\
\hline
\multirow{2}{*}{GPT-4.1-mini}
  & RL  & 0.4467 {\scriptsize$\pm$ 0.0075} & 0.4167 {\scriptsize$\pm$ 0.0149} & \textbf{0.4317 {\scriptsize$\pm$ 0.0107}} \\
  & SFT & 0.4483 {\scriptsize$\pm$ 0.0121} & 0.4250 {\scriptsize$\pm$ 0.0126} & 0.4283 {\scriptsize$\pm$ 0.0069} \\
\hline

\multirow{2}{*}{Gemini-2.0-Flash}
  & RL  & 0.4433 {\scriptsize$\pm$ 0.0047} & 0.4584 {\scriptsize$\pm$ 0.0075} & \textbf{0.4700} {\scriptsize$\pm$ 0.0100} \\
  & SFT & 0.4410 {\scriptsize$\pm$ 0.0068} & 0.4452 {\scriptsize$\pm$ 0.0085} & 0.4561 {\scriptsize$\pm$ 0.0094} \\
\hline

\end{tabular}%
}
\end{table*}

\subsection{Prompts}
\label{app:prompts}

\subsubsubsection{Agent Evaluation - Action Option Generation Prompt}

\begin{lstlisting}
You are an expert in Office Application automation and graphical user interfaces with accessibility support.

You will be provided with the following inputs:

1. **Current screenshot**: An image of the current state of an Office Application.
2. **Annotated current screenshot**: The same screenshot annotated with numeric markers corresponding to accessibility elements.2. **Annotated current screenshot**: The same screenshot annotated with numeric markers corresponding to accessibility elements.
3. **Accessibility (a11y) information**: This includes a list of control element labels and the textual name of the currently active Office Application.
4. **Task instruction**: A description of the action or goal to be completed.
5. **Supported actions**: A list of all actions that can be performed in this environment.

The annotated screenshot contains numbers that correspond directly to entries in the accessibility information. Each number identifies a specific UI control element, allowing you to reliably locate and reference interface components.

The accessibility information contains control labels that correspond to UI control elements in the current screenshot, allowing you to locate and reference specific interface components.

Your objective is to generate multiple diverse and plausible next actions that could be taken to accomplish the given task instruction, based on the current screenshot of the Office Application, the annotated current screenshot, and the available accessibility information. The desired number of options will be specified in the user instructions or input.

Use all the provided information-including the current screenshot, annotated screenshot, accessibility data, task instruction, and supported actions-to reason about the next appropriate actions accurately.


**IMPORTANT: When possible, prioritize using control_label over coordinate for actions. Control labels refer to unique identifiers provided by the accessibility (a11y) system for UI elements (such as buttons, text fields, or menu items). These identifiers are more reliable and accessible than raw screen coordinates, which may vary across layouts, resolutions, or UI states.**

For each candidate action, explain your reasoning process-describe how you analyze the current screenshot and the annotated screenshot, interpret the accessibility information, understand the current UI state, and determine what action could be taken next to move toward completing the task instruction.

Then, output the next actions in JSON format as a JSON array. Each element of the array must be an object with the keys "thoughts" and "tool_call". Both fields MUST be present in every element.

Please think very hard and carefully about the current state and the task instruction before making a decision, and output your reasoning in each element's "thoughts" field as detailed as possible, including:
- Your analysis of the current screenshot and annotated screenshot
- Your interpretation of the accessibility information
- How you identified the target control using control labels or visual elements
- The reason for your tool call and argument selection
- Your assessment of task progress based on the current state

The "tool_call" field in each element should contain:
- "function": str, The function/action type to execute
- "args": Dict, The arguments/parameters for the function
- "status": str, The status after performing this action (either "CONTINUE" or "FINISH")

For click operations, prioritize control_label over coordinate:
```json
{
  "thoughts": "The screenshot shows an Excel spreadsheet. From the accessibility information, there is a Save button with control_label=15. Using the control_label provides a more reliable interaction than estimating screen coordinates.",
  "tool_call": {
    "function": "click",
    "args": {"control_label": 15, "coordinate": null, "button": "left"},
    "status": "CONTINUE"
  }
}
````

If control_label is not available, fall back to coordinate:

```json
{
  "thoughts": "The target UI element does not have a corresponding control_label in the accessibility information. Based on the visual analysis of the annotated screenshot, I estimate the appropriate screen coordinates to interact with it.",
  "tool_call": {
    "function": "click",
    "args": {"control_label": null, "coordinate": [150, 30], "button": "left"},
    "status": "CONTINUE"
  }
}
```

For type operations, prioritize control_label over coordinate:

```json
{
  "thoughts": "The accessibility information indicates a text input field with control_label=8. Typing via the control_label ensures the correct field is targeted.",
  "tool_call": {
    "function": "type",
    "args": {"control_label": 8, "coordinate": null, "keys": "Hello World", "clear_current_text": true},
    "status": "CONTINUE"
  }
}
```

For drag operations, coordinates are required:

```json
{
  "thoughts": "This task requires dragging an element from one location to another. Drag actions require explicit start and end coordinates to describe the spatial movement.",
  "tool_call": {
    "function": "drag",
    "args": {"start_coordinate": [100, 100], "end_coordinate": [200, 200], "button": "left"},
    "status": "CONTINUE"
  }
}
```

If you think the task is finished, output status as "FINISH":

```json
{
  "thoughts": "Based on the current state of the Office Application and the task instruction, no further actions are required.",
  "tool_call": {
    "function": "",
    "args": {},
    "status": "FINISH"
  }
}
```

Only **ONE** action should be taken per option. You may output multiple options (as requested by num_options), but each option must correspond to exactly one action. If the task instruction could apply to multiple elements, choose the most relevant one for each option based on the current screenshot and accessibility information, and ensure the options are diverse.

Your response MUST be a valid JSON array with no additional text outside the JSON structure.

Task instruction:
{instruction}

Accessibility Information:
{a11y}

Supported actions:
{actions}

The current screenshot and the annotated current screenshot are provided as images.

Please analyze the current state using both visual information and accessibility data to generate {num_options} diverse and plausible next actions that could be taken to move toward completing the task instruction.

Please provide your reasoning and the next actions below in JSON array format without any additional text.
\end{lstlisting}

\subsubsection{Agent Evaluation - Action Selection Prompt}

\begin{lstlisting}

You are an AI assistant tasked with selecting the best next action in an Office application (e.g., Microsoft Word, Excel, or PowerPoint) based on a GUI screenshot and several candidate action outcomes.

The task instruction is:
[Instruction]

The current screenshot is: See the image below.

[Current State Screenshot]

You are given a list of action options, each with:
- action: the structured action command to be executed
- predicted_state_image: a visual prediction (mockup) showing the expected GUI state after executing the action

The `predicted_state` fields are simulations. Use them as references, but **ignore diffusion artifacts** (e.g., garbled text, noisy details). If a prediction is unreliable or inconsistent with the task, you may ignore it.

### Analysis Instructions:
1. **Analyze Every Option:** specific `predicted_state` simulations (images/text) are provided for every action. You must look at all of them first.
2. **When to use the World Model:**
   - **Uncertainty:** Use the predicted image when you are uncertain based purely on the action name or your internal knowledge.
   - **Discovery:** Use the predicted image if it clearly shows a better way of advancing the goal that you didn't initially think of.
   - **Efficiency:** Use the predicted image if it shows a way to skip intermediary steps and advance to the goal faster.
3. **Decide:** Combine your internal knowledge with these visual signals to pick the single best action.


Please return your final answer as a JSON object with the following format:
{
  "action_idx": <index of selected action>,
  "thought": "<brief explanation of why this action was selected>"
}
Do NOT include anything else outside the JSON object.
Here are the candidate action options:

Action Option 1:
 - Action:
 - Predicted State Image: see below.
[Predicted Image 1]

Action Option 2:
 - Action:
 - Predicted State Image: see below.
[Predicted Image 2]

Action Option 3:
 - Action: 
 - Predicted State Image: see below.
[Predicted Image 3]

Action Option 4:
 - Action: 
 - Predicted State Image: see below.
[Predicted Image 4]

Action Option 5:
 - Action: 
 - Predicted State Image: see below.
[Predicted Image 5]

Now, analyze all options and return only the selected 'action_idx' and a short 'thought' in JSON format. 'action_idx' MUST be an integer between 1 and 5 (inclusive). Do NOT output any text outside the JSON object.

\end{lstlisting}

\subsubsection{Textual State Transition Prompt}

\begin{lstlisting}
Your task is to predict and describe the likely appearance of the Next UI Screenshot, with a strong emphasis on how the UI would change compared to the current state, and where these changes would be located within the interface.

You are acting as a World Model assistant for Office applications such as {app_name}. Your output will be used by a text-to-image model to accurately reproduce the predicted Next UI Screenshot, while clearly encoding the UI transition from the previous state.

You are provided with:
- A screenshot of the current Office UI, which defines the baseline state
- A structured user action, provided for contextual grounding
- A function-level description of the GUI action and its arguments, provided for semantic reference

Important:
- The Next UI Screenshot is not available; you must generate a plausible prediction of what it would look like.
- Use the current screenshot to understand the baseline state.
- The user action and GUI description provide reference context for what changes are expected.
- Do not speculate beyond what would reasonably follow from the current UI and action.

---

Reference Information:

Current UI Screenshot:  
{image}

Action:  
{action}

GUI Action Description:  
{gui_description}

---

In your response, follow this structure:

1. Start by stating which Office application this is (e.g., "This is Microsoft PowerPoint.").
2. Briefly summarize the user interaction that would lead from the current UI to this predicted state, using the action only as context.
3. Predict and describe the Next UI Screenshot in a single coherent paragraph, explicitly highlighting how it would differ from the original screenshot, and specifying where those changes would likely occur.

When describing the UI, organize the description in a top-down order when applicable. Only describe changed parts; for unchanged elements, state explicitly (e.g., "Ribbon unchanged"):

- Title Bar (document name, window state, changes if any)
- Ribbon (active tab, visible groups, detailed controls and icons)
- Main Editing Area / Canvas (content, layout, selection state, unchanged or changed elements; emphasize position and alignment, e.g., center-aligned)
- Sidebar / Pane (opened, closed, or updated panels)
- Navigation Area (slide thumbnails, focus changes)
- Status Bar (zoom level, mode indicators)
- Dropdown / Popout (anchored to a specific control or cursor location, with its relative position and size explicitly described)

Explicitly indicate predicted changes and their locations using clear language, such as:
- "In the Ribbon, the 'Insert' tab is expected to become active..."
- "In the Main Editing Area, the text will likely change to 'Quarterly Report'..."
- "A new panel labeled 'Design Ideas' is likely to appear on the right sidebar..."

All visible text in the UI should be enclosed in double quotes (e.g., "Home", "File", "New Slide").

Ensure that your description:
- Clearly encodes the transition from the current UI to the next UI state
- Specifies where changes are likely to occur in the UI
- Uses terminology and layout conventions consistent with {app_name}

Do not include reasoning, internal thoughts, or references to the images as separate entities. Do not answer in bullet points.
Output a single paragraph of vivid, precise visual description suitable for text-to-image generation.
\end{lstlisting}

\subsubsection{Textual State Transition Evaluation - LLM-as-a-Judge Prompt}
\label{sec:llm_judge_prompt}
\begin{lstlisting}
You are an impartial LLM-as-a-Judge. Your task is to grade a model prediction (PRED) against the ground truth (GT) for describing the ``Next UI Screenshot'' of an Office application (e.g., Microsoft Word).

You MUST evaluate the following aspects independently:
1) App name
2) User action
3) Next-frame prediction:
   3.1) Title Bar
   3.2) Ribbon
   3.3) Main Editing Area / Canvas
   3.4) Sidebar / Pane
   3.5) Navigation Area
   3.6) Status Bar

Scoring rule for EACH aspect (use ONLY these values):
- 0   = completely incorrect / contradicts GT / missing when GT contains it
- 0.5 = partially correct: some key elements match, but has notable omissions or inaccuracies
- 1   = fully correct: matches GT on the key elements with no meaningful errors

Critical evaluation guidelines:
- Use GT as the single source of truth.
- Judge content fidelity, not writing quality.
- Be strict about factual UI elements (active tab name, document title, zoom %, panes open/closed, specific text edits).
- Penalize hallucinations: if PRED adds UI changes or elements not supported by GT, deduct in the relevant aspect(s).
- If GT does NOT mention a sub-area (e.g., Navigation Area), then:
  - If PRED also does not mention it -> score 1 (no contradiction).
  - If PRED claims a specific change/state that GT does not support -> score 0.5 or 0 depending on how strong/incorrect it is.
- When scoring 0.5 vs 1, treat the following as ``key elements'':
  - Title Bar: document name, saved/unsaved indicator, window state if mentioned
  - Ribbon: active tab, visible groups, important controls/menus if mentioned
  - Dropdown / Popout: presence, anchor, relative position, size, and visible content
  - Main Editing Area: the actual document text changes, formatting (bold/center-aligned), cursor/selection state, layout
  - Sidebar/Pane: which pane is open, its content list/state
  - Navigation Area: thumbnails/outline focus changes if present
  - Status Bar: page number, zoom, mode toggles (Track Changes, etc.)

Output format requirements:
- Output ONLY valid JSON.
- No markdown, no extra text.
- Include per-aspect scores.


Return JSON with exactly this structure:
{{
  "scores": {{
    "app_name": <0|0.5|1>,
    "user_action": <0|0.5|1>,
    "title_bar": <0|0.5|1>,
    "ribbon": <0|0.5|1>,
    "main_editing_area": <0|0.5|1>,
    "sidebar_pane": <0|0.5|1>,
    "navigation_area": <0|0.5|1>,
    "status_bar": <0|0.5|1>
  }},
  "notes": {{
    "app_name": "<one short sentence rationale>",
    "user_action": "<one short sentence rationale>",
    "title_bar": "<one short sentence rationale>",
    "ribbon": "<one short sentence rationale>",
    "main_editing_area": "<one short sentence rationale>",
    "sidebar_pane": "<one short sentence rationale>",
    "navigation_area": "<one short sentence rationale>",
    "status_bar": "<one short sentence rationale>"
  }}
}}

Now perform the evaluation.

PRED:
<<<
{PRED}
>>>

GT:
<<<
{GT}
>>>
\end{lstlisting}

\end{document}